\documentclass[aps,prb,10pt,superscriptaddress,twocolumn,twoside,nofootinbib]{revtex4-2}

\usepackage{amsmath,amsfonts,amssymb,amsthm,mathtools}

\usepackage[english]{babel}
\usepackage[utf8]{inputenc}
\usepackage[T1]{fontenc}

\usepackage{latexsym}
\usepackage[pdftex]{graphicx}
\usepackage{epstopdf}

\usepackage{subfigure}

\usepackage{enumitem}

\usepackage{hyperref}
\usepackage[all]{hypcap}

\hypersetup{
	pdftitle={},
	pdfauthor={},
	colorlinks=true,
	linkcolor=black,
	citecolor=black,
	filecolor=black,
	urlcolor=blue,
	bookmarksopen,
	bookmarksopenlevel=1,
}

\newcommand{\cB}{\mathcal{B}}
\newcommand{\cC}{\mathcal{C}}

\newcommand{\cF}{\mathcal{F}}

\newcommand{\cH}{\mathcal{H}}

\newcommand{\cQ}{\mathcal{Q}}
\newcommand{\cR}{\mathcal{R}}

\let\a=\alpha          \let\d=\delta

          \let\ph=\varphi

\newcommand{\ee}{\mathrm{e}}
\newcommand{\ii}{\mathrm{i}}
\newcommand{\dd}{\mathrm{d}}

\def\vF{v_{0}}

\def\io{\infty}
\def\eps{\epsilon}

\def\tq{\tilde{q}}

\def\sgn{\operatorname{sgn}}

\newcommand{\pdag}{^{\vphantom{\dagger}}}
\newcommand{\ppr}{^{\vphantom{\prime}}}
\newcommand{\xxa}{\stackrel{\scriptscriptstyle \times}{\scriptscriptstyle \times}\!}
\newcommand{\xxe}{\!\stackrel{\scriptscriptstyle \times}{\scriptscriptstyle \times}}
\newcommand{\fermionWick}[1]{\left. :\! \hspace{-0.5pt} #1 \hspace{-0.5pt} \!: \right.}
\newcommand{\bosonWick}[1]{\left. \xxa #1 \xxe \right.}
\newcommand{\anyonWick}[2]{N_{#2}[ #1 ] }

\theoremstyle{definition}

\theoremstyle{remark}

\begin{document}


\title{%
Approaching off-diagonal long-range order for 1+1-dimensional relativistic anyons%
}%

\author{%
Luca Fresta%
}%
\email[]{luca.fresta@math.uzh.ch }%
\affiliation{%
Institute of Mathematics, University of Zurich,
Winterthurerstrasse 190, 8057 Z{\"u}rich, Switzerland%
}%
\affiliation{%
Department of Mathematics, University of T{\"u}bingen,
Auf der Morgenstelle 10, 72076 T{\"u}bingen, Germany%
}%

\author{%
Per Moosavi%
}%
\email[]{pmoosavi@phys.ethz.ch}%
\affiliation{%
Institute for Theoretical Physics, ETH Zurich,
Wolfgang-Pauli-Strasse 27, 8093 Z{\"u}rich, Switzerland%
}%

\date{%
September 10, 2021%
}%


\begin{abstract}
We construct and study relativistic anyons in 1+1 dimensions generalizing well-known models of Dirac fermions. First, a model of free anyons is constructed and then extended in two ways: (i) by adding density-density interactions, as in the Luttinger model, and (ii) by coupling the free anyons to a U(1)-gauge field, as in the Schwinger model. Second, physical properties of these extensions are studied. By investigating off-diagonal long-range order (ODLRO) at zero temperature, we show that anyonic statistics allows one to get arbitrarily close to ODLRO but that this possibility is destroyed by the gauge coupling. The latter is due to a nonzero effective mass generated by gauge invariance, which we show also implies the presence of screening, independently of the anyonic statistics.
\end{abstract}


\maketitle


\section{Introduction}
\label{Sec:Introduction}


\renewcommand*{\thefootnote}{\fnsymbol{footnote}}


A simple way to describe anyons is as particles that satisfy exchange relations
with a phase interpolating between bosons and fermions.
They are often studied in 2+1 dimensions [\citealp{LeMy}, \citealp{GoMeSh1}, \citealp{GoMeSh2}, \citealp{Wilc}]\footnote[3]{Refs.\ \cite{GoMeSh1, GoMeSh2} added after publication of this paper; cf.\ also \cite{BLSW}.} and were recently experimentally demonstrated \cite{BKBMBBPCDGJF}, but such exchange relations are also possible in 1+1 dimensions; see, e.g., \cite{StWi, Iso, Poly, CaLa, Kund, PGL, BGO, BGH, WRDK, YMSY, PTT, LGDGG, CMT, SPC}.
For instance, anyonic fields can be constructed using vertex operators in 1+1-dimensional conformal field theory, and chiral versions have been proposed to describe edge currents associated to the fractional quantum Hall effect; see, e.g., \cite{Wen1, Wen2, Stern}. 
However, nonchiral such theories have received little to no attention, even though they are prime examples for studying the interplay between anyonic statistics and interactions by exact analytical means.

On the experimental side, there are recent proposals for realizing 1+1-dimensional anyons using ultracold atoms in optical lattices based on, e.g., Raman-assisted hopping \cite{KLMR, GrSa}, external drives \cite{SSE}, or multicolor modulation \cite{GCS}.
These mechanisms lead to occupation-number-dependent couplings between neighboring sites, which, following a generalized Jordan-Wigner transformation, yield an emergent anyonic description.
Other recent proposals involve impurity-based approaches \cite{LAS, YGLRLS} with the anyons appearing at the impurities.
We will not make reference to any specific underlying experimental setup but rather directly study the emergent (abelian) anyonic systems, here akin to nonchiral versions of the edge descriptions associated to the fractional quantum Hall effect.
One motivation is that these models are amenable to exact analytical computations, unlike \cite{KLMR, GrSa, SSE}, which instead rely on numerical techniques, or various approximations used in other related works, such as \cite{CMT}.
It would be interesting to explore if one can experimentally realize such anyonic systems using impurities as mentioned above, but this is beyond the scope of this paper.

To be more specific, in this paper we construct and analytically study nonchiral models of relativistic anyons in 1+1 dimensions that generalize well-known exactly solvable models of massless Dirac fermions.
First, we construct a model of free anyons, and then extend it in two ways:
(i) by adding density-density interactions, as in the Luttinger \cite{Tom, Lut, MaLi} and the Thirring \cite{Thirr} models, and
(ii) by coupling the free anyons to a $\mathrm{U}(1)$-gauge field, as in the Schwinger model \cite{Schw1, Schw2}.
Second, as applications, we study physical properties of these extensions.
Specifically, the possibility to observe off-diagonal long-range order (ODLRO) \cite{Yang, CCGOR} and screening effects \cite{LoSw, CKS2, CKS}.

The presence of ODLRO is characteristic of condensation according to the Penrose-Onsager criterion \cite{PeOn}, and the main object for studying ODLRO is the one-body density matrix (1BDM) \cite{Yang, CCGOR}.
In our case, the 1BDM can essentially be expressed as a 2$\times$2-matrix,
\begin{equation}
\label{1BDM}
\hat{\rho}(x, x')
= \left(
		\begin{matrix}
			\bigl\langle \psi^+_{+}(x) \psi^-_{+}(x') \bigr\rangle
			& \bigl\langle \psi^+_{+}(x) \psi^-_{-}(x') \bigr\rangle \\
			\bigl\langle \psi^+_{-}(x) \psi^-_{+}(x') \bigl\rangle
			& \bigl\langle \psi^+_{-}(x) \psi^-_{-}(x') \bigr\rangle
		\end{matrix}
	\right),
\end{equation}
whose elements are the equal-time two-point correlation functions of anyonic fields
$\psi^{-}_{r}(x)$ and $\psi^{+}_{r}(x) = \psi^{-}_{r}(x)^{\dagger}$
for $x \in [-L/2, L/2]$, where $L$ is the length of the system, and $r = +$$(-)$ denoting right (left) movers.
Here, we will exclusively consider ground-state correlation functions.
The fields satisfy
\begin{subequations}
\label{psi_psi_exch_rel_formal}
\begin{align}
\psi^{q}_{\pm}(x) \psi^{q'}_{\pm}(x')
& = \ee^{\mp\ii \pi qq' \alpha \sgn(x-x') }
		\psi^{q'}_{\pm}(x') \psi^{q}_{\pm}(x),
		\\
\psi^{q}_{\pm}(x) \psi^{q'}_{\mp}(x')
& = \ee^{\mp\ii \pi qq' \alpha}
		\psi^{q'}_{\mp}(x') \psi^{q}_{\pm}(x)
\end{align}
\end{subequations}
for $q, q' = \pm$ and $x \neq x'$, where $\alpha \in \mathbb{R}^{+}$ is called the \emph{statistics parameter}.
As we will see, our construction of the anyonic fields results in that $\alpha$ cannot be strictly zero, but it is possible to get arbitrarily close.

To check if ODLRO is present, one can look at the decay of the correlation functions for large $|x - x'|$, or equivalently, how their small Fourier modes scale with the system size \cite{PiSt, Note0}.
If the decay is algebraic, then the occupation number $\bar{n}_{k_{\mathrm{min}}}$ of the smallest mode $k_{\mathrm{min}}$ in the ground state will be shown to scale as
\begin{equation}
\label{n_propto_L}
\bar{n}_{k_{\mathrm{min}}} \propto L^{\cC}
\end{equation}
for some exponent $\cC$ \cite{CMT, CDMT}.
Since $L$ is extensive, $\cC = 1$ corresponds to ODLRO and thus true macroscopic condensation, i.e., the many-body phenomenon where a quantum state is macroscopically populated \cite{PiSt}.
However, even if $\cC$ is only arbitrarily close but not equal to $1$, one could argue that a majority of particles would still be in the smallest Fourier mode for any experimental system since experiments naturally are of finite size.
In this sense, there could still be an experimentally visible condensation effect, referred to as \emph{mesoscopic condensation}, at zero temperature \cite{CMT, CDMT}.

The free-anyon model that we will construct is formally given by the Hamiltonian
\begin{equation}
\label{H_0_formal}
H_{0}
= \sum_{r = \pm} \int_{-L/2}^{L/2} \dd x\,
		\anyonWick{
			\psi^+_{r}(x) (-\ii r \vF \partial_{x}) \psi^-_{r}(x)
		}{}
\end{equation}
(we use units where $\hbar = \epsilon_{0} = 1$),
where $\vF$ is the (bare) propagation velocity and $\anyonWick{\cdots}{}$ indicates a normal-ordering prescription for products of two anyonic fields (see Appendix~\ref{App:Anyon_normal_ordering} for details).
As mentioned, this generalizes free massless Dirac fermions in 1+1 dimensions by using fields $\psi^{\pm}_{r}(x)$ satisfying \eqref{psi_psi_exch_rel_formal} and by replacing the usual fermion Wick ordering by $\anyonWick{\cdots}{}$.

The first extension that we will consider is the \emph{anyonic Luttinger} (AL) \emph{model}.
Similar to the usual fermionic ($\alpha = 1$) model, it describes interacting right- and left-moving anyons with couplings between the anyon densities $\anyonWick{ \psi^+_{\pm}(x) \psi^-_{\pm}(x) }{}$.
For the AL model, we will show by exact analytical computations that the exponent $\cC$ in \eqref{n_propto_L} is
\begin{equation}
\label{cC_AL}
\cC_{\mathrm{AL}} = 1 - \alpha \frac{K^2 + 1}{2K},
\end{equation}
where $K > 0$ is the so-called Luttinger parameter and encodes the essential effects of the interactions [see \eqref{vK_g2_g4}]:
Here, $K = 1$ corresponds to noninteracting anyons and $K \neq 1$ to interacting ones.
We stress that $\cC_{\mathrm{AL}} < 1$ since $\alpha > 0$ in our construction, which means that true macroscopic condensation is not possible.
However, $\cC_{\mathrm{AL}}$ can take values arbitrarily close to $1$, which suggests that mesoscopic condensation is a possibility.

\begin{figure}[!htbp]

\centering

\subfigure[]{
\includegraphics[scale=1, trim=0 0 0 0, clip=true]{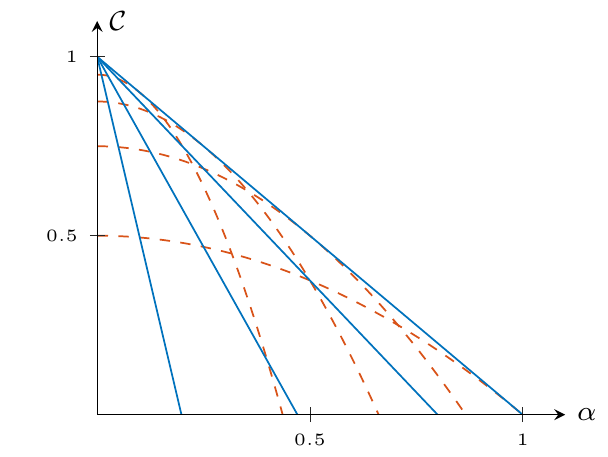}
\label{Fig:C_vs_a}
}

\vspace{-10pt}

\subfigure[]{
\includegraphics[scale=1, trim=0 0 0 0, clip=true]{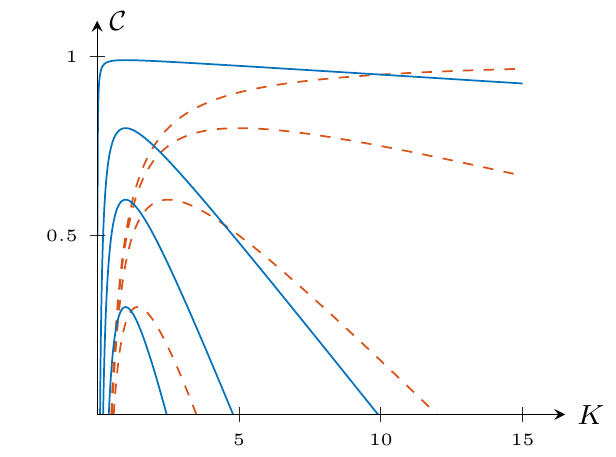}
\label{Fig:C_vs_K}
}

\vspace{-10pt}

\caption{%
Plots of $\mathcal{C}$ as a function \subref{Fig:C_vs_a} of $\alpha$ for $K = 1$, $2$, $4$, $10$ (right to left) and \subref{Fig:C_vs_K} of $K$ for $\alpha = 0.01$, $0.2$, $0.4$, $0.7$ (top to bottom) for the AL model in \eqref{cC_AL} (blue lines) and the ALL model in \eqref{cC_ALL} (red dashed lines).%
}%

\label{Fig:C_vs_aK}

\end{figure}

ODLRO was recently studied in \cite{CMT} for the anyonic generalization of the Lieb-Liniger (ALL) model.
The corresponding exponent $\cC$ in \eqref{n_propto_L} was found to be
\begin{equation}
\label{cC_ALL}
\cC_{\mathrm{ALL}}
= 1 - \frac{1}{2K} - \frac{K \alpha^2}{2},
\end{equation}
where $K$ describes the interaction in the ALL model within Luttinger-liquid theory \cite{Note1}.
As for AL anyons, mesoscopic condensation is seen to be possible for ALL anyons \cite{Note2}.
To compare these two results, we plot \eqref{cC_AL} and \eqref{cC_ALL} as functions of $\alpha$ and $K$ in Fig.~\ref{Fig:C_vs_aK}.

The second extension that we will consider is the \emph{anyonic Schwinger} (AS) \emph{model}.
As for the usual fermionic ($\alpha = 1$) model, this is obtained from \eqref{H_0_formal} by coupling the anyons to a $\mathrm{U}(1)$-gauge field via minimal coupling and by adding gauge dynamics.
We will show that $\hat{\rho}(x, x')$ in \eqref{1BDM} for the AS model decays exponentially with $|x - x'|$ due to a nonzero effective mass generated by gauge invariance.
This destroys any chance of observing ODLRO, even in the ground state.
We will also show that the same mass implies the presence of screening in the AS model, which we recall is a well-known property of the usual Schwinger model \cite{LoSw, CKS2, CKS}.
More concretely, one signature of screening is that the ``particle-antiparticle'' potential $V(d)$, here defined as the energy of two opposite external charges separated by a distance $d$ in the thermodynamic limit $L \to \io$, saturates to a constant as $d \to \io$ \cite{CKS2, AAR}.
For the AS model, we will show that
\begin{equation}
\label{particle_antiparticle_potential}
V_{\mathrm{AS}}(d)
= \frac{e_{0}^2}{2m\vF} \bigl( 1 - \ee^{-d m\vF} \bigr),
\end{equation}
where $m = e_{0}/\sqrt{\pi}\vF^{3/2}$ is the effective mass and $e_{0}$ is the elementary charge \cite{Note3}, which clearly tends to the constant $e_{0}^2/2m\vF$ as $d \to \io$.
This shows that the AS model exhibits screening in the above sense independently of the anyonic statistics.

We mention that screening was recently studied for a version of the fermionic Schwinger model with a gauge field in two or three spatial dimensions and shown to always be present even then \cite{BDT}.
Lastly, we remark that there are recent advances toward experimental realizations of the Schwinger model using ultracold atoms in optical lattices \cite{Cirac16, Nature16, LerMazz}.

The rest of this paper is organized as follows.
Our free anyonic model is constructed in Sec.~\ref{Sec:Free_anyons}.
The above mentioned extensions are presented in Sec.~\ref{Sec:Extensions}.
ODLRO and screening for these extensions are studied in Sec.~\ref{Sec:Applications}.
Concluding remarks are given in Sec.~\ref{Sec:Concluding_remarks}.


\section{Free anyons}
\label{Sec:Free_anyons}


Models of 1+1-dimensional anyons have been considered as emergent descriptions in strongly correlated fermionic systems.
A primary example arises in connection with fractional quantum Hall states on a compact geometry, see, e.g., \cite{Wen1}, where the edge excitations are shown to have fractional charge [cf.\ \eqref{anyonic_charge}].
A more recent example can be found in \cite{LAS}, where anyons appear at magnetic impurities in a chiral multi-channel Kondo problem.

Even though the anyons that we will construct here would emerge in the context of strongly correlated systems, it is justifiable to call them free since mathematically they are a simple extension of free massless Dirac fermions in 1+1 dimensions.
The approach that we will present is based on vertex operators in the spirit of \cite{CaLa, LaMo1}, which is well known in the context of edge excitations associated to the fractional quantum Hall effect; see, e.g., \cite{Wen1, ElSch}.
Since we are interested in nonchiral models, we will construct two flavors of anyons, one right- and one left-moving, and prescribe exchange relations between them.


\subsection{Hilbert space}
\label{Sec:Free_anyons:Hilbert_space}


We begin by introducing the Hilbert space $\cF$ on which our anyonic fields will be defined.
As will become clear, in our construction, we need two parameters $\alpha_{0} \in \mathbb{R}^{+}$ and $\alpha \in \mathbb{R}^{+}$ with the condition
\begin{equation}
\label{n_a}
n_{\a} = \sqrt{\alpha/\alpha_{0}} \in \mathbb{Z}^{+}.
\end{equation}
Let $\rho\pdag_{r}(p)^{\dagger} = \rho\pdag_{r}(-p)$ and $R_{r}^{\vphantom{-1}\dagger} = R_{r}^{-1\vphantom{\dagger}}$ for $r = \pm$ and $p \in (2\pi/L) \mathbb{Z}$ be operators satisfying
\begin{subequations}
\begin{align}
\bigl[ \rho_{r\ppr}(p), \rho_{r'}(p') \bigr]
& = r \frac{Lp}{2\pi} \delta_{r, r'} \delta_{p+p', 0},
		\label{rho_rho_CCR} \\
\bigl[ \rho_{r\ppr}(p), R_{r'} \bigr]
& = r \sqrt{\alpha_{0}} \delta_{r, r'}\delta_{p, 0} R_{r\ppr},
		\label{rho_R_CCR}
\end{align}
\end{subequations}
and
\begin{subequations}
\begin{gather}
\rho_{r}(rp) |\Psi_{0}\rangle = 0
\quad
\forall p \geq 0, \\
\langle\Psi_{0}| R^{q_{+}}_{+}R^{-q_{-}}_{-} |\Psi_{0}\rangle
= \delta_{q_{+}, 0} \delta_{q_{-}, 0}
\end{gather}
\end{subequations}
for $q_{\pm} \in \mathbb{Z}$,
which also defines the vacuum $|\Psi_{0}\rangle$.
Operators with different $r = \pm$ commute, except for the $R_{\pm}$-operators for which we have a freedom to impose exchange relations.
For reasons that will become clear, we choose
\begin{equation}
\label{R_exchange_relations}
R\pdag_{\pm}R\pdag_{\mp}
= \ee^{\pm\ii\pi \alpha_{0}} R\pdag_{\mp}R\pdag_{\pm}.
\end{equation}
Evidently, (odd) even $\alpha_{0}$ corresponds to (anti-) commutation relations.

We refer to $\rho_{\pm}(p)$ as density operators, $\sqrt{\alpha_{0}} Q_{\pm} = \rho_{\pm}(0)$ as zero modes, and $R_{\pm}$ as Klein factors.
For later reference, we note that $R_{r}^{r n_{\a}}$ and $R_{r}^{-r n_{\a}}$ can be interpreted as charge raising and lowering operators, respectively, and $\sqrt{\alpha_{0}/\alpha} Q_{r}$ as chiral anyon number operators.
Moreover, if we introduce the elementary charge $e_{0}$, then $e_{0} \sqrt{\alpha_{0}}Q_{r}$ can be interpreted as chiral charge operators.
Indeed, it follows from \eqref{rho_R_CCR} that
$\bigl[ \sqrt{\alpha_{0}}Q_{r}, R_{r}^{\pm r n_{\a}} \bigr]
= \pm \sqrt{\alpha} R_{r}^{\pm r n_{\a}}$.
This means that $R_{r}^{\pm r n_{\a}}$ change the charge in units of
\begin{equation}
\label{anyonic_charge}
e = \sqrt{\alpha} e_{0}.
\end{equation}
As we will see, $e$ can be interpreted as anyonic (or fractional) charge.
We note that the reason for the condition in \eqref{n_a} is that only integer powers of $R_{\pm}$ are well-defined.
Moreover, the same condition implies that $Q_{\pm}$ have integer spectrum.

Finally, by using the density operators and the Klein factors, it is possible to construct from $|\Psi_{0}\rangle$ a set of orthonormal states $|\Psi_{0, \mathbf{m}}\rangle$ for $\mathbf{m} = \{ ( m(p) )_{p \neq 0}, q_+, q_- \}$,
$m(p) \in \mathbb{N} = \{ 0, 1, \ldots \}$ and $q_{\pm} \in \mathbb{Z}$, where at most finitely many of the $m(p)$ are nonzero, and from these construct $\cF$; see Appendix~\ref{App:Free_Hilbert_space} for further details.
For clarity, $|\Psi_{0, \mathbf{0}}\rangle = |\Psi_{0}\rangle$ and is the ground state of $H_{0}$ in \eqref{H_0_formal} [made precise in \eqref{H_0}].


\subsection{Anyonic fields}
\label{Sec:Free_anyons:Anyonic_fields}


Given $\eps > 0$, we define the regularized fields
\begin{widetext}
\begin{equation}
\label{Reg_anyonic_fields}
\psi^{-}_{r}(x; \eps)
= L^{-\alpha/2}
	\! \bosonWick{
		R^{-r n_{\a}}_{r}
		\exp \Biggl(
			\ii r \sqrt{\alpha} \frac{2\pi}{L}
				\biggl[
					x \sqrt{\alpha_{0}} Q_{r}
						+ \sum_{p \neq 0}\frac{1}{\ii p} \rho_{r}(p) \ee^{\ii px - \eps |p|/2}
				\biggr]
		\Biggr)
	},
\quad
\psi^{+}_{r}(x; \eps) = \psi^{-}_{r}(x; \eps)^{\dagger}
\end{equation}
for $r = \pm$ and $x \in [-L/2, L/2]$, where $\bosonWick{ \cdots }$ indicates boson Wick ordering; see \cite{LaMo1} for details.
From here it is clear that we must require $\alpha > 0$ for this construction to be meaningful.
Using the tools in \cite{LaMo1}, one can show that [cf.\ \eqref{psi_psi_exch_rel_formal}]
\begin{subequations}
\label{Reg_anyon_exchange_relations}
\begin{align}
\psi^{q}_{\pm}(x; \eps) \psi^{q'}_{\pm}(x'; \eps')
& = \ee^{\mp\ii \pi qq' \alpha \sgn(x-x'; \eps + \eps')}
		\psi^{q'}_{\pm}(x'; \eps') \psi^{q}_{\pm}(x; \eps),
		\label{Reg_anyon_exchange_relations_1} \\
\psi^{q}_{\pm}(x; \eps) \psi^{q'}_{\mp}(x'; \eps')
& = \ee^{\mp\ii \pi qq' \alpha}
		\psi^{q'}_{\mp}(x'; \eps') \psi^{q}_{\pm}(x; \eps)
		\label{Reg_anyon_exchange_relations_2}
\end{align}
\end{subequations}
\end{widetext}
for $x, x' \in [-L/2, L/2]$, $q, q' = \pm$, and $\eps, \eps' > 0$,
where
$\sgn(x; \eps)$
is a regularized sign function \cite{Note4}.
We see from \eqref{Reg_anyon_exchange_relations} that (odd) even $\a$ corresponds to (anti-) commutation relations, but from \eqref{anyonic_charge} that, e.g., $\a = 1$ and $\a = 3$ are still different.
[Note that \eqref{Reg_anyon_exchange_relations_2} follows from \eqref{Reg_anyonic_fields} and our choice in \eqref{R_exchange_relations}.]

The definition in \eqref{Reg_anyonic_fields} is similar to the construction of regularized fermionic fields based on vertex operators reviewed in \cite{LaMo1}.
Here, our fields are anyonic in the sense that they satisfy the generalized exchange relations in \eqref{Reg_anyon_exchange_relations}.
Moreover, it follows from Sec.~\ref{Sec:Free_anyons:Hilbert_space} that the fields in \eqref{Reg_anyonic_fields} change the charge in units of $e$ in \eqref{anyonic_charge}.
This motivates interpreting $e$ as anyonic charge.
In general, as explained below [see \eqref{Reg_anyonic_fields_BC}], the boundary conditions for the anyonic fields are twisted and depend on the charge sector of the Hilbert space.

Using the tools in \cite{LaMo1}, we can compute exact nonequal-time correlation functions for the anyonic fields
$\psi^{\pm}_{r}(x, t; \eps)
= \ee^{\ii H_{0} t} \psi^{\pm}_{r}(x; \eps) \ee^{-\ii H_{0} t}$.
The results for the ground-state two-point correlation functions are
\begin{align}
& \lim_{\eps, \eps' \to 0^{+}}
	\langle\Psi_{0}|
		\psi^{+}_{r}(x, t; \eps) \psi^{-}_{r'}(x', t'; \eps')
	|\Psi_{0}\rangle \nonumber \\
& \quad = G_{r, r'}(x - x'; t - t')
	\label{Free_psipsi_2pcf_def}
\end{align}
with
\begin{equation}
\label{Free_psipsi_2pcf_result}
\lim_{L \to \io} G_{r, r'}(x; t) \\
= \frac{\delta_{r, r'}}{(2\pi)^{\alpha}}
	\biggl( \frac{r\ii}{x - r\vF t + \ii r0^{+}} \biggr)^{\alpha}
\end{equation}
in the thermodynamic limit.


\subsection{Some remarks}
\label{Sec:Free_anyons:Some_remarks}


Before continuing, a number of remarks are in order:
\begin{enumerate}[wide, labelindent=1em, nosep, label={\textnormal{(\roman*)}}, ref={\textnormal{(\roman*)}}]

\item
\label{Some_remarks:2nd_set_of_fields}
One can introduce a second set of fields given by \eqref{Reg_anyonic_fields} with $\sqrt{\alpha}$ replaced by $1/\sqrt{\alpha}$ and $n_{\alpha} = 1$; see, e.g., \cite{Iso, ElSch, AtLa}.
However, in this paper, these fields will not be considered.

\item
\label{Some_remarks:Restricted_Hilbert_space}
If the fields in Remark~\ref{Some_remarks:2nd_set_of_fields} are not included, then $\alpha_{0}$ can be absorbed by
(a) defining $\cR_{r}^{\pm} = R_{r}^{\pm n_{\a}}$ and $\cQ_{r} = \sqrt{\alpha_{0}} Q_{r}$, and
(b) restricting $q_{r}$ in the sequence $\textbf{m}$ to integer multiples of $n_{\a}$ in \eqref{n_a}, i.e., $q_{r} = \tq_{r} n_{\a}$ for $\tq_{r} \in \mathbb{Z}$.
The reason is that the full Hilbert space $\cF$ is equivalent to a fermionic Fock space through the boson-fermion correspondence, see, e.g., \cite{LaMo1}, while (b) defines a restricted Hilbert space $\cF_{\mathrm{res}}$ of $\cF$.
In other words, (a) and (b) ``remove'' this connection with fermions, and the result is closer to the approach of constructing anyons starting from compactified bosons; see, e.g., \cite{Iso, Note5}.
In particular,
$\bigl[ \cQ_{r}, \cR_{r}^{\pm r \tq_{r}} \bigr]
= \pm \sqrt{\alpha} \tq_{r} R_{r}^{\pm r \tq_{r}}$,
meaning that $\cQ_{\pm}/\sqrt{\alpha}$ have integer spectrum in $\cF_{\mathrm{res}}$.

\item
\label{Some_remarks:Boundary_conditions}
The restricted Hilbert space $\cF_{\mathrm{res}}$ above can be decomposed as
$\cF_{\mathrm{res}}
= \oplus_{(\tq_{+}, \tq_{-}) \in \mathbb{Z}^2} \cF_{\mathrm{res}}^{(\tq_{+}, \tq_{-})}$
into charge sectors $\cF_{\mathrm{res}}^{(\tq_{+}, \tq_{-})}$.
In each of these, the anyonic fields can be shown to obey twisted boundary conditions
\begin{equation}
\label{Reg_anyonic_fields_BC}
\psi^{\pm}_{r}(x+L; \eps)
= \ee^{\mp \ii r \pi \phi(\tq_{r})} \psi^{\pm}_{r}(x; \eps)
\end{equation}
with
$\phi(\tq_{r}) = \alpha (2\tq_{r} - 1)$.
In other words, the anyons know about the charge of the system.
For later reference, we define the Fourier transforms of our fields as
$\hat{\psi}_{r}^{\pm}(k; \eps)
= (2\pi)^{-1/2} \int_{-L/2}^{L/2} \dd x\, \psi^{\pm}_{r}(x; \eps) \ee^{\pm \ii rkx}$ for
$k \in (2\pi/L)[\mathbb{Z} + \phi(\tq_{r})/2]$.
(Note the $r$ in $\ee^{\pm \ii rkx}$.)
We stress that these sets of momenta depend on the charge sector through $\phi(\tq_{r})$ but otherwise do not depend on $r$.

\item
\label{Some_remarks:Rational_alpha}
On the restricted Hilbert space, if $\alpha$ is rational, then \eqref{Reg_anyonic_fields_BC} can be rephrased as
\begin{equation}
\psi^{\pm}_{r}(x+wL;\eps)
= \psi^{\pm}_{r}(x;\eps)
\end{equation}
for some integer $w$.

\end{enumerate}
%


\section{Extensions}
\label{Sec:Extensions}


Since our anyons constructed in the previous section generalize free massless Dirac fermions, it is natural to consider the following two extensions with many-body interactions:
(i) The anyonic Luttinger (AL) model, where we add density-density interactions.
(ii) The anyonic Schwinger (AS) model, where we instead couple the free anyons to a $\mathrm{U}(1)$-gauge field.
For these extensions, following the discussion of edge descriptions associated to the fractional quantum Hall effect in Secs.~\ref{Sec:Introduction} and~\ref{Sec:Free_anyons}, one physical picture one could have in mind is that of two such edges with oppositely-moving anyons in close proximity.


\subsection{The anyonic Luttinger model}
\label{Sec:Extensions:AL_model}


Our first extension is formally given by
\begin{equation}
\label{H_AL}
H_{\mathrm{AL}} = H_{0} + H_{\mathrm{int}} - E_{\mathrm{AL}, 0}
\end{equation}
with $H_{0}$ in \eqref{H_0_formal} and
\begin{align}
H_{\mathrm{int}}
& = \sum_{r, r' = \pm}
		\int_{-L/2}^{L/2} \dd x\,
		\frac{\pi\vF}{2} 
		\bigl( g_{2} \delta_{r, -r'} + g_{4} \delta_{r, r'} \bigl)
		\nonumber \\
& \quad \; \times
		\anyonWick{ \psi^+_{r\ppr}(x) \psi^-_{r\ppr}(x) }{}
		\anyonWick{ \psi^+_{r'}(x) \psi^-_{r'}(x) }{},
		\label{H_int_formal}
\end{align}
where $g_{2}$ and $g_{4}$ are dimensionless couplings satisfying $|g_{2}| < 2 + g_{4}$ and we subtracted the energy $E_{\mathrm{AL}, 0}$ of the interacting ground state $|\Psi_{\mathrm{AL}}\rangle$.
The couplings describe density-density interactions between ($g_{2}$) and among ($g_{4}$) the right- and left-moving anyons, similar to the usual fermionic ($\alpha = 1$) Luttinger model; see, e.g., \cite{Voit, SCP}.
See Appendix~\ref{App:Solution_of_AL_model} for details (including how to make the AL model precise and to solve it).

To study the AL model, it is convenient to introduce the renormalized velocity and the Luttinger parameter
\begin{align}
v
& = \vF \sqrt{ (1 + g_{4}/2)^2 - (g_{2}/2)^2 }, \nonumber \\
K
& = \sqrt{ (2 + g_{4} - g_{2})/(2 + g_{4} + g_{2}) }, \label{vK_g2_g4}
\end{align}
respectively \cite{Voit, SCP}.
The first is the propagation velocity in the interacting model, and the second will be seen to encode the essential effects of the interactions.

It is well known that point-like interactions as in \eqref{H_int_formal} lead to ultraviolet divergencies.
To handle these requires:
(i) Additive renormalization of the Hamiltonian,
i.e., the subtraction of $E_{\mathrm{AL}, 0}$ above.
(ii) Multiplicative renormalization of the anyonic fields in \eqref{Reg_anyonic_fields}; see Appendix~\ref{App:Solution_of_AL_model} for details.
For ease of notation, we will here use $\psi^\pm_{r}(x; \eps)$ to also denote the renormalized fields.

As before for free anyons, using the tools in \cite{LaMo1}, one can compute nonequal-time correlations functions for the renormalized anyonic fields
$\psi^\pm_{r}(x, t; \eps)
= \ee^{\ii H_{\mathrm{AL}}t} \psi^\pm_{r}(x, t; \eps) \ee^{-\ii H_{\mathrm{AL}}t}$.
The results for the ground-state two-point correlation functions are
\begin{align}
& \lim_{\eps, \eps' \to 0^{+}}
	\langle\Psi_{\mathrm{AL}}|
		\psi^+_{r\ppr}(x, t; \eps)
		\psi^-_{r'}(x', t'; \eps')
	|\Psi_{\mathrm{AL}}\rangle \nonumber \\
& \quad = G^{\mathrm{AL}}_{r, r'}(x - x'; t - t')
	\label{AL_psipsi_2pcf_def}
\end{align}
with
\begin{align}
& \lim_{L \to \io}
	G^{\mathrm{AL}}_{r, r'}(x; t) \nonumber \\
& \;\;
	= \frac{\delta_{r, r'}}{(2\pi)^{\alpha}}
		\biggl(
			\frac{ \ii}{x - vt + \ii 0^{+}}
		\biggr)^{2\Delta_{r}^{+}}
		\biggl(
			\frac{-\ii}{x + vt - \ii 0^{+}}
		\biggr)^{2\Delta_{r}^{-}}
		\label{AL_psipsi_2pcf_result}
\end{align}
in the thermodynamic limit, where
\begin{equation}
\label{Delta_r_pm}
\Delta_{r}^{\pm} = \alpha \frac{(K \pm r)^2}{8K}.
\end{equation}
For $\alpha = 1$, the latter can be recognized as the conformal weights for the fields in the usual Luttinger model.
Thus, for anyons they are generalized by the factor $\alpha$.

As a remark, we stress that the correlation functions in \eqref{AL_psipsi_2pcf_result} correspond to those in Klaiber's solutions \cite{Kla} of the Thirring model (see Eq.~(VII.3) in \cite{Kla}).
More precisely, Klaiber computed all $N$-point correlation functions for that model and showed that, in general, they belong to a two-parameter family consisting of a statistics parameter $\lambda$ and a coupling constant $g$.
These can be shown to correspond to our parameters $\alpha$ and $K$ via
$\lambda = \pi(\alpha - 1)$ and $g^2 = \alpha \pi^2 (K-1)^2/K$.
Thus, in principle, already \cite{Kla} allowed for 1+1-dimensional anyons (see Eq.~(VI.2) in \cite{Kla}).


\subsection{The anyonic Schwinger model}
\label{Sec:Extensions:AS_model}


In our second extension, we couple the free anyonic model in \eqref{H_0_formal} to a $\mathrm{U}(1)$-gauge field.
In one spatial dimension, the electromagnetic tensor consists of one nontrivial component only, i.e., $F_{\mu \mu'} = \epsilon_{\mu \mu'} E/c$, where $E = E(x)$ is the electric field, $\epsilon_{\mu \mu'}$ is the Levi-Civita symbol ($\epsilon_{0 1} = 1$), and $c$ is the speed of light.
The gauge degrees of freedom consist of $E(x)$ with its conjugated field $A(x)$ \cite{Note6}.
The overall Hilbert space of the theory is $\cH = \cF \otimes \cF_{\mathrm{gauge}}$ with $\cF$ introduced in Sec.~\ref{Sec:Free_anyons:Hilbert_space} and $\cF_{\mathrm{gauge}}$ in
Appendix~\ref{App:Gauge_field}.
However, as discussed in Appendices~\ref{App:Gauge_field:Gauge_transformations} and~\ref{App:Gauge_field:Gauss_law}, $\cH$ is redundant due to gauge invariance.
We will therefore ``restrict'' to a physical Hilbert space $\cH_{\mathrm{phys}}$ of gauge-invariant states, which will correspond to working in the Coulomb gauge.

The Hamiltonian of the model is obtained by making $H_{0}$ gauge invariant and by adding gauge dynamics.
The former requires the usual substitution for gauge covariant derivatives, i.e., $\ii \partial_{x} \to \ii \partial_{x} - e A(x)$, where $e$ is the anyonic charge in \eqref{anyonic_charge}.
However, this is not enough due to the singular nature of the fields:
One must also replace $\anyonWick{\cdots}{}$ by
a gauge-invariant normal ordering $\anyonWick{\cdots}{\mathrm{AS}}$ of products of the fields; see Appendix~\ref{App:Gauge_field:Anyon_normal_ordering}. 
Using this, we can formally write
\begin{align}
H_{\mathrm{AS}}
& = \frac{1}{2} \int_{-L/2}^{L/2} \dd x\, \! \bosonWick{ E(x)^{2} } \!
	+ \sum_{r = \pm} \int_{-L/2}^{L/2} \dd x\, \nonumber \\
& \quad \times
		\anyonWick{
			\psi^{+}_{r}(x) r\vF [-\ii \partial_{x} + e A(x)] \psi^{-}_{r}(x)
		}{\mathrm{AS}} - E_{\mathrm{AS},0}, \label{H_AS}
\end{align}
where the last term is an additive renormalization, with $E_{\mathrm{AS}, 0}$ denoting the ground-state energy in the absence of external charges.
The normal ordering of the first term in \eqref{H_AS} describing the gauge dynamics is
explained in Appendix~\ref{App:Solution_of_AS_model}.

The model is not defined only by the Hamiltonian since physical solutions also have to obey Gauss's law.
In the absence of external charges, the latter can formally be written as
\begin{equation}
G(x) = 0,
\end{equation}
where $G(x)$ is the so-called Gauss's-law operator, whose Fourier components are given by
\begin{equation}
\hat G(p)
= - \ii p \hat{E}(p) + \frac{e}{\sqrt{\alpha}} \rho(p)
\quad
\forall p,
\end{equation}
where $\rho(p) = \rho_{+}(p) + \rho_{-}(p)$.
In the physical Hilbert space, Gauss's law is an actual identity and corresponds to fixing a trivial representation of small gauge transformations on $\cH_{\mathrm{phys}}$; see Appendices~\ref{App:Gauge_field:Gauge_transformations} and~\ref{App:Gauge_field:Gauss_law}.
For completeness, we remark that $\cH_{\mathrm{phys}}$ and, in particular, the ground state of the model depend on the density of external charges if present; see Appendix~\ref{App:Solution_of_AS_model}.

In the AS model, as for the usual Schwinger model, global chiral symmetry is spontaneously broken and the ground state is infinitely degenerate:
We denote it by
$|\Psi_{\mathrm{AS}}(\theta)\rangle$ for $\theta \in [0,2\pi)$; see Appendix~\ref{App:Gauge_field:Gauss_law}.
Solving the model, see Appendix~\ref{App:Solution_of_AS_model}, shows that it has the relativistic dispersion relation
\begin{equation}
\label{omega_AS}
\omega(p) = \sqrt{p^2 \vF^2 + m^2 \vF^4}
\end{equation}
with the effective mass
\begin{equation}
\label{effective_mass}
m
= \frac{e}{\sqrt{\pi \alpha} \vF^{3/2}}
= \frac{e_{0}}{\sqrt{\pi} \vF^{3/2}}.
\end{equation}

The gauge-invariant ground-state two-point correlation functions for the AS model in the Coulomb gauge are defined as
\begin{multline}
\label{AS_psipsi_2pcf_def}
G^{\mathrm{AS}}_{\theta; r, r'}(x, x'; t, t')
= \lim_{\eps, \eps' \to 0^{+}} 
	\langle\Psi_{\mathrm{AS}}(\theta)| 
		\psi_{r}^{+}(x, t; \eps)
		\ee^{-\ii w_{\eps\ppr}(x, t)} \\
\times
		\ee^{\ii w_{\eps'}(x', t')}
		\psi_{r'}^{-}(x', t'; \eps')
	|\Psi_{\mathrm{AS}}(\theta)\rangle,
\end{multline}
using the Wilson-line operator
\begin{equation}
\label{Wilson-line_operator}
w_{\eps}(x, t) = e \int_{-L/2}^{L/2} \dd y\, \cB_{\epsilon}(x, y) A(y, t),
\end{equation}
where $\cB_{\eps}(x, y) = x/L - \sum_{p \neq 0} (1/\ii pL) \ee^{\ii p(x-y) - \eps|p|/2}$
and $A(y, t)$ is the time-evolved gauge potential.
As for free anyons in Sec.~\ref{Sec:Free_anyons:Anyonic_fields} and the AL model in Sec.~\ref{Sec:Extensions:AL_model}, such correlation functions can be computed by exact analytical means:
The explicit expression for \eqref{AS_psipsi_2pcf_def} is given in \eqref{AS_psipsi_2pcf_result} and \eqref{AS_psipsi_2pcf_result_parts} in Appendix~\ref{App:2pcf_in_the_AS_model}.


\section{Applications}
\label{Sec:Applications}


In this section we discuss certain important physical properties of the AL and AS models introduced in Sec.~\ref{Sec:Extensions}, specifically off-diagonal long-range order at zero temperature and screening effects.


\subsection{Approaching ODLRO for the anyonic Luttinger model}
\label{Sec:Applications:AL_model}


To study the possibility of observing ODLRO for the AL model in the ground state, we consider the occupation number $\bar{n}_{k}$ of Fourier mode $k$ for finite $L$, which can be expressed as
\begin{equation}
\label{n_k_L}
\bar{n}_{k}
= \sum_{r = \pm} \int_{-L/2}^{L/2} \dd x\,
	\lim_{\eps \to 0^{+}}
	\langle\Psi_{\mathrm{AL}}|
		\psi^+_{r}(x; \eps)
		\psi^-_{r}(0; \eps)
	|\Psi_{\mathrm{AL}}\rangle
	\ee^{-\ii rkx}
\end{equation}
for $k \in (2\pi/L)(\mathbb{Z} - \alpha/2)$ [cf.\ Remark~\ref{Some_remarks:Boundary_conditions} in Sec.~\ref{Sec:Free_anyons:Some_remarks}].
We recall that this is directly related to the Fourier transform of the 1BDM \cite{PeOn, PiSt}; see Sec.~\ref{Sec:Introduction}.
Such occupation numbers were computed also by Mattis and Lieb in \cite{MaLi} (and used to show that sufficiently strong interactions in the usual Luttinger model can eliminate the Fermi surface).
Note that momenta $k$ in our model are measured relative to the vacuum, which for fermions is referred to as the filled Dirac sea.

The correlation functions in \eqref{AL_psipsi_2pcf_result} decay algebraically as
$|x|^{-\alpha (K^2 + 1)/2K}$ for large $|x|$.
For our system, this decay can be shown to be true also for finite but large $L$ requiring only that $L \gg |x| \gg 0$.
Thus, in Fourier space, the occupation number $\bar{n}_{k_{\mathrm{min}}}$ of the smallest mode $k_{\mathrm{min}}$ must scale as
\begin{equation}
\label{n_k_min_L_bar}
\bar{n}_{k_{\mathrm{min}}}
\propto
(\Delta k)^{\alpha (K^2 + 1)/(2K) - 1},
\end{equation}
where $\Delta k = 2\pi/L$ is the separation between points in Fourier space for finite $L$ and $k_{\mathrm{min}} = -\pi(\alpha - 2n_{0})/L$ for some $n_{0} \in \mathbb{Z}$ such that $\alpha - 2n_{0} \in (-1, 1]$.
The result in \eqref{n_k_min_L_bar} implies:%

\emph{%
In the anyonic Luttinger model, the occupation number $\bar{n}_{k_{\mathrm{min}}}$ of the smallest mode $k_{\mathrm{min}}$ scales as $\bar{n}_{k_{\mathrm{min}}} \propto L^{\cC_{\mathrm{AL}}}$ for large $L$ with $\cC_{\mathrm{AL}}$ in \eqref{cC_AL} for $\alpha > 0$ and $K > 0$.%
}%

Recall from Sec.~\ref{Sec:Introduction} that, if $\cC_{\mathrm{AL}} = 1$, then we would have ODLRO and thus macroscopic condensation, while $\cC_{\mathrm{AL}}$ close but not exactly equal to $1$ corresponds to mesoscopic condensation.
Moreover, if $\cC_{\mathrm{AL}} < 0$,
then $\bar{n}_{k_{\mathrm{min}}} \to 0$ as $L \to \io$.
Thus, from the above result and \eqref{cC_AL}, we draw the following conclusions:%
\begin{enumerate}[wide, labelindent=1em, nosep, label={\textnormal{(\roman*)}}, ref={\textnormal{(\roman*)}}]

\item
For $\alpha \geq 1$, there can never be macro- or mesoscopic condensation since $(K^2 + 1)/2K$ for $K > 0$ has a global minimum at $K = 1$ (i.e., no interactions).
As expected, this includes fermions ($\alpha = 1$).

\item
For $0 < \alpha < 1$, mesoscopic condensation is possible, even with interactions, by making $\alpha$ arbitrarily small.
However, since $\alpha$ cannot be strictly zero, macroscopic condensation is never possible.

\end{enumerate}

Lastly, we note that, for a given $\alpha$, it is always possible to destroy any mesoscopic condensation by changing the interactions [cf.\ the formula for $K$ in \eqref{vK_g2_g4}].


\subsection{Absence of ODLRO and presence of screening in the anyonic Schwinger model}
\label{Sec:Applications:AS_model}


To study ODLRO for the AS model we consider the 1BDM directly.
As explained in Sec.~\ref{Sec:Introduction}, the elements of the latter consist of the two-point correlation functions $G^{\mathrm{AS}}_{\theta; r, r'}(x, x'; 0, 0)$ in \eqref{AS_psipsi_2pcf_def}; see \eqref{AS_psipsi_2pcf_result} and \eqref{AS_psipsi_2pcf_result_parts} in Appendix~\ref{App:2pcf_in_the_AS_model} for the explicit expression.
There, we also state and prove a number of properties for these correlation functions, including that
$G^{\mathrm{AS}}_{\theta; r, r'}(x, x'; 0, 0)$ decays exponentially in $|x - x'|$ due to the nonzero mass in \eqref{effective_mass}.
Specifically, we show that $\lim_{L \to \io} G^{\mathrm{AS}}_{\theta; r, r'}(x, x'; 0, 0)$ for $|x - x'| \gg 1$ decays like
\begin{equation}
\exp \bigl( - C m\vF \alpha |x - x'| \bigr)
\end{equation}
for some constant $C > 0$.
This implies:%

\emph{%
In the anyonic Schwinger model, ODLRO is always absent independently of $\alpha > 0$.%
}%

Another property we prove in Appendix~\ref{App:2pcf_in_the_AS_model} is that global chiral symmetry is recovered as the mass vanishes in the sense that
$\lim_{m \to 0^{+}} \lim_{L \to \io} G^{\mathrm{AS}}_{\theta; \pm, \mp}(x, x'; t, t') = 0$. 
We also show that $G^{\mathrm{AS}}_{\theta; r, r'}(x, x'; t, t')$ has a peculiar behavior in the thermodynamic limit:
If $x \neq x'$, then
$\lim_{L \to \io} G^{\mathrm{AS}}_{\theta; r, r'}(x, x'; t, t') = 0$
unless $t - t' \in (2\pi/m\vF^2) \mathbb{Z}$.
Moreover, the limit is approached in an exponential way that depends on $t - t'$, and thus removing this vanishing contribution is not simply a matter of a multiplicative renormalization.

By studying the correlation functions in \eqref{AS_psipsi_2pcf_def} it is also possible to probe the presence of screening; see, e.g., \cite{CKS}.
Another way to investigate this, which is also more direct, is to study the ``particle-antiparticle'' potential \cite{CKS2}.
This is defined as the difference between the ground-state energy of the model when two opposite external charges are present and the case when they are absent.
In one spatial dimension, one would naturally expect this energy difference to grow linearly with the distance between the two charges, unless screening occurs.
In the latter case, the external charges, regardless of their magnitude, are effectively surrounded by a cloud of opposite charges and therefore the potential saturates to a constant value for long distances.

As remarked in Sec.~\ref{Sec:Extensions:AS_model}, external charges do not modify the Hamiltonian as written in \eqref{H_AS}, but they do modify Gauss's law, which formally becomes
\begin{equation}
\label{Gauss_law_ext_charges}
G(x)
= - \frac{e}{\sqrt{\alpha}}\rho_{\mathrm{ext}}(x),
\end{equation}
where $\rho_{\mathrm{ext}}(x)$ is the density of external charges.
Consequently, the ground state as well as the effective Hilbert space $\cH_{\mathrm{phys}}$ are modified; see Appendix~\ref{App:Gauge_field:Gauss_law}.
We are interested in the following special case:%
\begin{equation}
\label{ext_charge_deltas}
\rho_{\mathrm{ext}}(x) = \delta(x - d/2) - \delta(x + d/2),
\end{equation}
which describes two opposite point charges separated by a distance $d < L/2$.
Denote by $|\Psi_{\mathrm{AS}}(\theta, d)\rangle$ the corresponding ground state.
The ``particle-antiparticle'' potential is then
$V_{\mathrm{AS}}(d)
= \lim_{L \to \io}
	\langle\Psi_{\mathrm{AS}}(\theta, d)| H |\Psi_{\mathrm{AS}}(\theta, d)\rangle
	- \langle\Psi_{\mathrm{AS}}(\theta)| H |\Psi_{\mathrm{AS}}(\theta)\rangle$, which can be shown to give exactly the result in \eqref{particle_antiparticle_potential}; see Appendix~\ref{App:Solution_of_AS_model}.
Since $m \neq 0$ in \eqref{effective_mass} for any $\alpha \in \mathbb{R}^{+}$,
it follows that $V_{\mathrm{AS}}(d)$ saturates to a constant, which implies:%

\emph{%
In the anyonic Schwinger model, there is always screening independently of $\alpha > 0$.%
}%
 
Note that this result is due to the same nonzero mass as is responsible for the exponential decay of $G^{\mathrm{AS}}_{\theta; r, r'}(x, x'; 0, 0)$ in $|x - x'|$.
Indeed, it follows from \eqref{particle_antiparticle_potential} that
$\lim_{m \to 0^{+}} V_{\mathrm{AS}}(d) \propto d$,
as expected for two oppositely charged particles in the absence of screening.


\section{Concluding remarks}
\label{Sec:Concluding_remarks}


We constructed a free model of 1+1-dimensional relativistic anyons and two different extensions that we called (i) the anyonic Luttinger (AL) and (ii) the anyonic Schwinger (AS) model.
In (i) we added density-density interactions, and in (ii) we instead coupled the free anyons to a $\mathrm{U}(1)$-gauge field.
We studied physical properties of these extensions, specifically the possibility to observe ODLRO and screening effects.
For the AL model we showed that one can get arbitrarily close to ODLRO, which we referred to as mesoscopic condensation, while ODLRO is always absent in the AS model.
The reason for the latter was shown to be related to the presence of screening, which we showed is always there, independently of the anyonic statistics.

Our results for the AL model were compared with recent results for an anyonic generalization of the Lieb-Liniger (ALL) model in \cite{CMT}.
In general, $\cC_{\mathrm{AL}}$ in \eqref{cC_AL} and $\cC_{\mathrm{ALL}}$ in \eqref{cC_ALL} depend differently on $K$ and $\alpha$.
However, it is interesting to note that for fermions ($\alpha = 1$) they agree for all $K > 0$.

In addition to the approach followed in this paper based on vertex operators with generalized (exchange) statistics, there is another approach to anyons based on Haldane's fractional exclusion statistics (FES) \cite{Hald, Wu}; see, e.g., \cite{Poly} for a review.
Simply put, FES measures how the available Hilbert space changes with the addition of one more particle.
Thus, a natural extension of this work would be to investigate the connections between our approach and that based on FES; cf., e.g., \cite{MuSh}.
To this end, one possible route motivates the inclusion of the second set of fields in Remark~\ref{Some_remarks:2nd_set_of_fields} in Sec.~\ref{Sec:Free_anyons:Some_remarks} as those are expected to be needed if one wants to construct the full Hilbert space using anyon creation and annihilation operators \cite{Iso, ElSch, AtLa}.

Lastly, we note that it would be interesting to study the more general anyonic model with both density-density interactions and coupling to a gauge field.
This would be an anyonic generalization of the fermionic model in \cite{GLR} and allow one to investigate the full interplay between anyonic statistics, interactions, and gauge coupling.
It would also be interesting to study possible connections between the observed possibility to get arbitrarily close to ODLRO for the AL model and questions about stability of matter for anyons \cite{DoSo1, DoSo2}.


\acknowledgments{%
We want to thank Gian Michele Graf, Edwin Langmann, Douglas Lundholm, Stefano Scopa, and Andrea Trombettoni for valuable discussions and helpful remarks.
P.M.\ gratefully acknowledges financial support from the Wenner-Gren Foundations (Grant No.\ WGF2019-0061).
L.F.\ is thankful for support from the Swiss National Science Foundation (via the grant ``Mathematical Aspects of Many-Body Quantum Systems'') and from the European Research Council (ERC) under the European Union's Horizon 2020 research and innovation programme (ERC Starting Grant MaMBoQ, Grant Agreement No.\ 802901).%
}%


\begin{appendix}


\section{Hilbert space and solution of the free-anyon model}
\label{App:Free_Hilbert_space}


To construct the Hilbert space $\cF$ in Sec.~\ref{Sec:Free_anyons:Hilbert_space}, we introduce the boson creation and annihilation operators
\begin{equation}
b^+_{p}
= \left( b^-_{p} \right)^{\dagger},
\quad
b^-_{p}
= \begin{cases}
		- \ii\sqrt{\frac{2\pi}{L|p|}} \rho_{+}(p) & \textnormal{if } p > 0, \\
		+ \ii\sqrt{\frac{2\pi}{L|p|}} \rho_{-}(p) & \textnormal{if } p < 0. \\
	\end{cases}
\label{b_pm_p}
\end{equation}
It follows that these satisfy
\begin{equation}
\bigl[ b^-_{p\ppr}, b^+_{p'} \bigr] = \d_{p, p'},
\quad
\bigl[ b^\pm_{p\ppr}, b^\pm_{p'} \bigr] = 0,
\quad
b^-_{p}|\Psi_{0}\rangle = 0
\quad
\forall p \neq 0.
\end{equation}
Consider the states
\begin{equation}
\label{Psi_0_bfm}
|\Psi_{0, \mathbf{m}}\rangle
= \left( \prod_{p \neq 0} \frac{\bigl( b_{p}^{+} \bigr)^{m(p)}}{\sqrt{m(p)!}} \right)
	R_+^{q_+} R_-^{-q_-} |\Psi_{0}\rangle
\end{equation}
for $\mathbf{m} = \{ ( m(p) )_{p \neq 0}, q_+, q_- \}$,
$m(p) \in \mathbb{N} = \{ 0, 1, \ldots \}$ and $q_{\pm} \in \mathbb{Z}$, where at most finitely many of the $m(p)$ are nonzero.
One can check that these states form an orthonormal basis, and $\cF$ can be constructed from the vector space of all finite linear combination of these states; see, e.g., \cite{LaMo1}.

The free-anyon Hamiltonian in \eqref{H_0_formal} can be made precise as
\begin{equation}
\label{H_0}
H_{0}
= \vF \sum_{r = \pm}
	\Biggl[
		\frac{\pi}{L} \alpha_{0} Q_{r}^2
		+ \sum_{p>0} \frac{2\pi}{L} \rho_{r}(-rp) \rho_{r}(rp)
	\Biggr];
\end{equation}
see Appendix~\ref{App:Anyon_normal_ordering}.
The ground state of $H_{0}$ is $|\Psi_{0}\rangle$, and $|\Psi_{0, \mathbf{m}}\rangle$
in \eqref{Psi_0_bfm} are its exact eigenstates with the corresponding exact eigenvalues
\begin{equation}
\label{cE_0_bfm}
E_{0, \mathbf{m}}
= \frac{\pi\vF}{L} \alpha_{0} \bigl( q_{+}^2 + q_{-}^2 \bigr)
	+ \sum_{p \neq 0} \vF |p| m(p),
\end{equation}
i.e., $H_{0} |\Psi_{0, \mathbf{m}}\rangle = E_{0, \mathbf{m}} |\Psi_{0, \mathbf{m}}\rangle$.


\section{Anyon normal ordering}
\label{App:Anyon_normal_ordering}


We introduce an anyon normal-ordering prescription $\anyonWick{\cdots}{}$ for bilinears in the anyonic fields inspired by operator product expansions \cite{Note7}.
Formally, we define
\begin{equation}
\label{anyon_normal_ordering}
\anyonWick{ \psi^{+}_{r}(x) \partial^{n}_{x} \psi^{-}_{r}(x) }{}
= \lim_{\eps \to 0^{+}} \lim_{y \to 0}
	\anyonWick{ \psi^{+}_{r}(x; \eps) \partial^{n}_{y} \psi^{-}_{r}(x+y; \eps) }{\eps}
\end{equation}
for $n = 0, 1, \dots$, where 
\begin{multline}
\label{anyon_normal_ordering_eps}
\anyonWick{
	\psi^{+}_{r}(x; \eps) \partial^{n}_{y} \psi^{-}_{r}(x+y; \eps)
}{\eps} \\
= \frac{r}{2\pi \ii} \frac{L^{\alpha}}{(n+1)!}
	\biggl( \frac{1}{\sqrt{\alpha}} \frac{\partial}{\partial y} \biggr)^{n+1}
	\! \bosonWick{ \psi^{+}_{r}(x; \eps) \psi^{-}_{r}(x+y; \eps) } \!
\end{multline}
for $y \neq 0$.
As we will see below, this normal-ordering prescription is useful because we can formally write
\begin{equation}
\rho_{\pm}(x)
= \anyonWick{ \psi^{+}_{\pm}(x) \psi^{-}_{\pm}(x) }{}
\end{equation}
and $H_{0}$ in \eqref{H_0} as in \eqref{H_0_formal}.
In particular, we emphasize that our prescription implies that $H_{0}$ in \eqref{H_0_formal} is independent of $\alpha$, in agreement with \eqref{H_0}.

To show these connections, using standard manipulations with vertex operators, we write
\begin{multline}
\! \bosonWick{ \psi^{+}_{r}(x; \eps)\psi^{-}_{r}(x+y; \eps) } \!
= L^{-\alpha} \! \xxa
		\exp \biggl(
			\ii r \sqrt{\alpha} \\
\times
			\frac{2\pi}{L}
			\biggl[
				y \rho_{r}(0)
				+ \sum_{ p \neq 0} \frac{\ee^{\ii p y} - 1}{\ii p}
					\rho_{r}(p) \ee^{\ii p x - \eps |p|/2}
			\biggr]
		\biggr)
	\xxe.
\end{multline}
Therefore, using \eqref{anyon_normal_ordering_eps}, we obtain
\begin{equation}
\lim_{y \to 0}
\anyonWick{ \psi^{+}_{r}(x; \eps) \psi^{-}_{r}(x+y; \eps) }{\eps}
= \sum_{p} \frac{1}{L} \rho_{r}(p) \ee^{\ii p x - \eps |p|/2},
\end{equation}
which, by comparing with \eqref{anyon_normal_ordering}, means that we can formally identify
$\anyonWick{ \psi_{r}^{+}(x)\psi_{r}^{-}(x) }{}$ with $\rho_{r}(x)$.
Likewise,
\begin{align}
& \lim_{y \to 0}
	\anyonWick{ \psi^{+}_{r}(x; \eps) \partial_{y} \psi^{-}_{r}(x+y; \eps) }{\eps}
	\nonumber \\
& \quad
	= \frac{1}{2\sqrt{\alpha}} \partial_{x}
		\biggl( \sum_{p} \frac{1}{L} \rho_{r}(p) \ee^{\ii p x - \eps |p|/2} \biggr)
		\nonumber \\
& \qquad
	+ \ii \pi r
		\! \bosonWick{
			\biggl( \sum_{p} \frac{1}{L} \rho_{r}(p) \ee^{\ii p x - \eps |p|/2} \biggr)^{2}
		},
\end{align}
and thus, using \eqref{anyon_normal_ordering}, we obtain
\begin{align}
& \int_{-L/2}^{L/2} \dd x\,
	\anyonWick{ \psi^{+}_{r}(x)(-\ii r \vF \partial_{x})\psi^{-}_{r}(x ) }{} \nonumber \\
& \quad
	= \frac{\pi\vF}{L} \rho_{r}(0)^2
		+ \sum_{p>0} \frac{2\pi\vF}{L} \rho_{r}(-rp) \rho_{r}(rp)
\end{align}
for $r = \pm$, which implies that the expression in \eqref{H_0_formal} formally corresponds to the one in \eqref{H_0}.

To conclude this appendix, we establish a connection with the usual ordering for fermions ($\alpha = 1$) based on point splitting:%
\begin{multline}
\label{fermionic_normal_ordering}
:\! \psi^{+}_{r}(x)\psi^{-}_{r}(x)\!: \; 
= \lim_{\eps \to 0^{+}} \lim_{y \to 0}
	\Big(\psi^{+}_{r}(x; \eps)\psi^{-}_{r}(x+y; \eps) \\
- \langle\Psi_{0}| \psi^{+}_{r}(x; \eps) \psi^{-}_{r}(x+y; \eps) |\Psi_{0}\rangle \Big).
\end{multline}
One can check that the expression in parentheses on the right-hand side is sufficiently smooth in $y$ and that
\begin{align} 
& \lim_{\eps \to 0^{+}} \lim_{y \to 0}
	\langle\Psi_{0}| \psi^{+}_{r}(x; \eps)\psi^{-}_{r}(x+y; \eps) |\Psi_{0}\rangle ^{-1}
= 0, \nonumber \\
& \lim_{\eps \to 0^{+}} \lim_{y \to 0} \partial_{y}
	\langle\Psi_{0}| \psi^{+}_{r}(x; \eps)\psi^{-}_{r}(x+y; \eps) |\Psi_{0}\rangle ^{-1}
= 2\pi \ii r.
\end{align}
Since
\begin{align}
& \! \bosonWick{ \psi^{+}_{r}(x; \eps)\psi^{-}_{r}(x+y; \eps) } \! \nonumber \\
& = \frac{L^{-\alpha}}
			{\langle\Psi_{0}| \psi^{+}_{r}(x; \eps)\psi^{-}_{r}(x+y; \eps) |\Psi_{0}\rangle}
		\psi^{+}_{r}(x; \eps)\psi^{-}_{r}(x+y; \eps),
\end{align}
we obtain $\anyonWick{ \psi^{+}_{r}(x)\psi^{-}_{r}(x) }{}
= \! \fermionWick{ \psi^{+}_{r}(x)\psi^{-}_{r}(x) }$ if $\alpha = 1$.


\section{Solution of the anyonic Luttinger model}
\label{App:Solution_of_AL_model}


We briefly summarize the solution of the AL model in Sec.~\ref{Sec:Extensions:AL_model} following the presentations in \cite{LaMo1, LLMM1}.

First, as for $H_{0}$ in \eqref{H_0}, the interaction term $H_{\mathrm{int}}$ in \eqref{H_int_formal} can be written as
\begin{align}
& H_{\mathrm{int}}
	= \frac{\vF}{2}
		\sum_{r = \pm}
		\biggl\{
			\frac{\pi}{L} \alpha_{0}
			\Bigl[
				g_{2} Q_{r}Q_{-r} + g_{4} Q_{r}^2
			\Bigr] \nonumber \\
&	\quad
		+ \sum_{p>0} \frac{2\pi}{L}
			\Bigl[
				g_{2} \rho_{r}(-p) \rho_{-r}(p) + g_{4} \rho_{r}(-p) \rho_{r}(p)
			\Bigr]
		\biggr\}. \label{H_int}
\end{align}
This can be seen using the anyon normal-ordering prescription in Appendix~\ref{App:Anyon_normal_ordering}.
As for $H_{0}$, we emphasize that our normal ordering implies that $H_{\mathrm{int}}$ in
\eqref{H_int_formal} does not depend on $\alpha$, in agreement with \eqref{H_int}.

Second, to make our discussion more precise, we introduce the nonlocal version of the AL model, which is obtained by replacing $\pi\vF$ in \eqref{H_int} by momentum-dependent interaction potentials $V_{2}(p)$ and $V_{4}(p)$ in each term proportional to $g_{2}$ and $g_{4}$, respectively.
The precise conditions on the interactions are
\begin{subequations}
\label{g_conditions}
\begin{gather}
V_{2,4}(p) = V_{2,4}(-p),
	\label{g_conditions_1} \\
\bigl| g_{2}V_{2}(p) \bigr| < 2\pi\vF + g_{4}V_{4}(p)
	\quad \forall p,
	\label{g_conditions_2} \\
\sum_{p > 0} \frac{ p \bigl[ g_{2}V_{2}(p) \bigr]^2 }
	{ 2\pi\vF \bigl[ 2\pi\vF + g_{4}V_{4}(p) \bigr] } < \io.
	\label{g_conditions_3}
\end{gather}
\end{subequations}

When the conditions in \eqref{g_conditions} are satisfied, the Hamiltonian $H_{\mathrm{AL}}$ can be diagonalized by a Bogoliubov transformation implemented by a unitary operator $\ee^{\ii S_{\mathrm{AL}}}$, where
\begin{equation}
\label{S_AL}
S_{\mathrm{AL}}
= \ii \sum_{p \neq 0} \frac{2\pi}{L} \frac{\ph_{\mathrm{AL}}(p)}{p} \rho_{+}(-p)\rho_{-}(p)
\end{equation}
with $\ph_{\mathrm{AL}}(p) = \ph_{\mathrm{AL}}(-p) \in \mathbb{R}$ given by
$\tanh 2\ph_{\mathrm{AL}}(p)
= -g_{2}V_{2}(p)/[2\pi\vF + g_{4}V_{4}(p)]$
for all $p \neq 0$.
The corresponding momentum-dependent propagation velocity and Luttinger parameter are
\begin{align}
v(p)
& = \vF
		\sqrt{
			\biggl[ 1 + \frac{g_{4}V_{4}(p)}{2\pi\vF} \biggr]^2
			- \biggl[ \frac{g_{2}V_{2}(p)}{2\pi\vF} \biggr]^2
		}, \nonumber \\
K(p)
& = \ee^{2\ph_{\mathrm{AL}}(p)}
	= \sqrt{
			\frac{ 2\pi\vF + g_{4}V_{4}(p) - g_{2}V_{2}(p) }
				{ 2\pi\vF + g_{4}V_{4}(p) + g_{2}V_{2}(p) }
		},
\end{align}
respectively.
For the Hamiltonian, we obtain
\begin{equation}
\label{H_AL_diagonalized}
\ee^{\ii S_{\mathrm{AL}}} H_{\mathrm{AL}} \ee^{-\ii S_{\mathrm{AL}}}
= H_{\mathrm{AL}}^{Q} + D_{\mathrm{AL}},
\end{equation}
where $H_{\mathrm{AL}}^{Q}$ is the zero-mode contribution to $H_{\mathrm{AL}}$ and $D_{\mathrm{AL}}$ is diagonal in $\rho_{r}(p \neq 0)$.
Moreover, for nonlocal interactions, the interacting ground state
$|\Psi_{\mathrm{AL}}\rangle$
can be written
$|\Psi_{\mathrm{AL}}\rangle = \ee^{-\ii S_{\mathrm{AL}}} |\Psi_{0}\rangle$,
and similarly for excited states using $|\Psi_{0, \mathbf{m}}\rangle$ in Appendix~\ref{App:Free_Hilbert_space}.
The corresponding ground-state energy
$E_{\mathrm{AL}, 0} = - \sum_{p >0} [\vF - v(p)] p$
subtracted in \eqref{H_AL} diverges in the limit of point-like interactions.

As mentioned, for point-like interactions, the Hamiltonian and the anyonic fields have to be renormalized.
The first renormalization is already taken care of by subtracting $E_{\mathrm{AL}, 0}$.
For the second, consider sharp cutoffs $V_{i}(p) = \pi \vF \Theta(\pi/a - |p|)$ for $a > 0$, where $\Theta(\cdot)$ is the Heaviside function.
Each anyonic field $\psi^\pm_{r}(x; \eps)$ gives rise to a factor
\begin{equation}
Z_{a, \eps}
= \exp
	\biggl(
		- \alpha \sum_{p > 0} \frac{2\pi}{Lp}
			\frac{[1 - K(p)]^2}{4K(p)} \ee^{-\eps p}
	\biggr)
\end{equation}
in any correlation function; see, e.g., \cite{LaMo1}.
Since
$Z_{a, \eps}
\xrightarrow{a \to 0^{+}} \bigl( 1 - \ee^{-2\pi\eps/L} \bigr)^{\alpha (1-K)^2/4K}
\xrightarrow{\eps/L \to 0^{+}} 0$, we must renormalize $\psi^\pm_{r}(x; \eps)$ by multiplying them by $Z_{a, \eps}^{-1}$.
To get meaningful results for large $L$, they must also be rescaled by
$\bigl( 2\pi\tilde{\ell}/L \bigr)^{\alpha (1-K)^2/4K}$,
where $\tilde{\ell} > 0$ is an arbitrary length.
Since
$\bigl( 2\pi\tilde{\ell}/L \bigr)^{\alpha (1-K)^2/4K} Z_{a, \eps}^{-1}
\to
\bigl( \ee^{\gamma} \pi\tilde{\ell}/a \bigr)^{\alpha (1-K)^2/4K} [1 + O(a/L))]$
as $\eps \to 0^{+}$, where $\gamma$ is the Euler-Mascheroni constant,
we more specifically renormalize $\psi^\pm_{r}(x; \eps)$ by multiplying them by
$( \ee^{\gamma} \pi\tilde{\ell}/a )^{\alpha (1-K)^2/4K}$
and set $\tilde{\ell} = 1$ as is common.


\section{Solution of the anyonic Schwinger model}
\label{App:Solution_of_AS_model}


Below we give key steps in the solution of the AS model in Sec.~\ref{Sec:Extensions:AS_model}.
These are then used to derive \eqref{particle_antiparticle_potential}.

First of all, the term  
$(1/2) \int_{-L/2}^{L/2} \dd x\, \! \bosonWick{E(x)^{2}}$ 
in \eqref{H_AS} really stands for
\begin{equation}
\frac{\hat{E}(0)^{2}}{2L}
+ \sum_{p \neq 0} \frac{1}{2L p^{2}}
	\! \xxa
		\biggl[ \frac{e}{\sqrt{\alpha}} \rho(-p) - \hat{G}(-p) \biggr]
		\biggl[ \frac{e}{\sqrt{\alpha}} \rho(p)  - \hat{G}(p)  \biggr]
	\xxe,
\end{equation}
where $\bosonWick{ \cdots }$ indicates the boson Wick ordering introduced in \eqref{Reg_anyonic_fields}.
Accordingly, the Hamiltonian of the model can be written as (cf.\ Appendix~\ref{App:Gauge_field:Anyon_normal_ordering})
\begin{multline}
\label{H_AS_Fourier_space}
H_{\mathrm{AS}}
= H_{\mathrm{AS}}^{Q}
	+ \sum_{p \neq 0} \frac{\pi\vF}{2L}
		\Biggl(
			\frac{\omega(p)^{2}}{\vF^2 p^{2}}
			\! \bosonWick{ \rho^{G}(-p) \rho^{G}(p)} \! \\
			+ \! \bosonWick{ \rho^{G}_{5}(-p) \rho^{G}_{5}(p) } \!
			+ \frac{\vF \hat{G}(-p)\hat{G}(p)}{\pi \omega(p)^{2}}
		\Biggr)
	- E_{\mathrm{AS}, 0}
\end{multline}
with the zero-mode contribution
\begin{equation}
H_{\mathrm{AS}}^{Q}
= \frac{\hat{E}(0)^{2}}{2L}
	+ \frac{\pi \vF \rho^{G}_{5}(0)^{2}}{2 L}
	- \frac{m\vF^2}{2}
	+ \frac{\vF^{2}\hat{G}(0)^{2}}{2 L \omega(0)},
\end{equation}
where
\begin{align}
\rho^{G}(p)
& = \rho^{G}_{+}(p) + \rho^{G}_{-}(p),
\quad
\rho^{G}_{5}(p)
= \rho^{G}_{+}(p) - \rho^{G}_{-}(p), \nonumber \\
\rho^{G}_{r}(p)
& = \rho_{r}(p)
		+ r \frac{e \hat{A}(p)}{\sqrt{\alpha}} 
		- \frac{e \hat{G}(p)}{2\pi \sqrt{\alpha} \omega(p)^{2}}
\end{align}
with
$\hat{A}(p) = (2\pi)^{-1} \int_{-L/2}^{L/2} \dd x\, A(x) \ee^{-\ii px}$.
Similar to the solution of the AL model in Appendix~\ref{App:Solution_of_AL_model}, we define
\begin{equation}
S_{\mathrm{AS}}
= \ii \sum_{p \neq 0} \frac{2\pi}{L}
	\frac{\varphi_{\mathrm{AS}}(p)}{p} \rho^{G}_{+}(-p)\rho^{G}_{-}(p)
\end{equation}
with $\varphi_{\mathrm{AS}}(p) = \varphi_{\mathrm{AS}}(-p) \in \mathbb{R}$ given by 
$\tanh 2 \varphi_{\mathrm{AS}}(p)
= m^{2}\vF^{2}/(2 p^{2} + m^{2}\vF^{2})$
for all $p \neq 0$.
It follows that
\begin{equation}
\label{H_AS_diagonalized}
\ee^{\ii S_{\mathrm{AS}}} H_{\mathrm{AS}} \ee^{-\ii S_{\mathrm{AS}}}
= H_{\mathrm{AS}}^{Q}
	+ D_{\mathrm{AS}}
	+ \sum_{p \neq 0} \frac{\vF^{2}\hat{G}(-p)\hat{G}(p)}{2L \omega(p)^{2}},
\end{equation}
where $D_{\mathrm{AS}}$ is diagonal in $\rho^{G}_{r}(p \neq 0)$.
The ground-state energy, subtracted in \eqref{H_AS}, is
$E_{\mathrm{AS}, 0} = - \sum_{p >0} [\omega(p) - p \vF]^{2}/(2p \vF)$
and diverges as $L \to \io$.
The zero-mode contribution requires a separate discussion which we omit
for conciseness.

Moreover, similarly to the free case, cf.\ Appendix~\ref{App:Free_Hilbert_space}, we can construct exact eigenstates
$|\Psi_{\mathrm{AS}, \mathbf{n}}(\theta)\rangle$ for
$\mathbf{n} = ( n(p) )_{p \in (2\pi/L) \mathbb{Z}}$, $n(p) \in \mathbb{N}$, where at most finitely many of the $n(p)$ are nonzero.
These satisfy
$H_{\mathrm{AS}} |\Psi_{\mathrm{AS}, \mathbf{n}}(\theta)\rangle
= E_{\mathrm{AS}, \mathbf{n}} |\Psi_{\mathrm{AS}, \mathbf{n}}(\theta)\rangle$
with the exact eigenvalues
\begin{equation}
E_{\mathrm{AS}, \mathbf{n}}
= \sum_{p} \omega(p) n(p),
\end{equation}
using $\omega(p)$ in \eqref{omega_AS}.

The result in \eqref{H_AS_diagonalized} can be used to derive \eqref{particle_antiparticle_potential} as follows.
In the presence of external charges, the operators $\hat{G}(p)$
[for $p \in (2\pi/L)\mathbb{Z}$] have a trivial action in $\cH_{\mathrm{phys}}$, see Appendix~\ref{App:Gauge_field:Gauss_law}, and it follows from \eqref{Gauss_law_ext_charges} and \eqref{H_AS_diagonalized} that the excitation energies are modified according to
\begin{align}
& H_{\mathrm{AS}} |\Psi_{\mathrm{AS}, \mathbf{n}}(\theta, \rho_{\mathrm{ext}})\rangle
\nonumber \\
& = \biggl[
			E_{\mathrm{\mathrm{AS}, \mathbf{n}}}
			+ \sum_{p}
				\frac{e^{2}\vF^{2} \rho_{\mathrm{ext}}(-p)\rho_{\mathrm{ext}}(p)}
					{2L \alpha \omega(p)^2}
		\biggr]
		|\Psi_{\mathrm{AS}, \mathbf{n}}(\theta, \rho_{\mathrm{ext}})\rangle,
\end{align}
where $\rho_{\mathrm{ext}}(p) = \int_{-L/2}^{L/2} \dd x\, \rho_{\mathrm{ext}}(x) \ee^{-\ii px}$.
From this, using that \eqref{ext_charge_deltas} implies
$\rho_{\mathrm{ext}}(p) = -2\ii \sin(pd/2)$,
we read out the ``particle-antiparticle'' potential to be
\begin{equation}
V_{\mathrm{AS}}(d)
= \frac{e^{2}_{0}}{\pi} \int_{-\io}^{\io} \dd p\,
	\frac{\sin^{2}(pd/2)}{p^{2} + m^{2}\vF^2},
\end{equation}
which gives the expression in \eqref{particle_antiparticle_potential}.


\section{Two-point correlation functions for the anyonic Schwinger model}
\label{App:2pcf_in_the_AS_model}


Using the tools in \cite{LaMo1} one can show that
\begin{align}
\lim_{L \to \io}
G^{\mathrm{AS}}_{\theta; r, r'}(x, x'; t, t')
& = \lim_{L \to \io}
		\ee^{\ii (r-r') \tilde{\theta}/2}
		\tilde{G}_{0; r, r'}(x \hspace{-0.5mm} - \hspace{-0.5mm} x'; t \hspace{-0.5mm} - \hspace{-0.5mm}t') \nonumber \\
& \quad \times
		R_{r, r'}(t - t')
		\ee^{\alpha K_{r, r'}(x - x'; t - t')}
		\label{AS_psipsi_2pcf_result}
\end{align}
with $\tilde{\theta} = \theta n_{\a}$ [cf.\ Remark~\ref{Some_remarks:Restricted_Hilbert_space} in Sec.~\ref{Sec:Free_anyons:Some_remarks}] and
\begin{widetext}
\begin{align}
\tilde{G}_{r, r'}(x; t)
& = L^{- \alpha \delta_{r, r'}}
		\ee^{- \ii r \pi \alpha \d_{r, r'} x/L}
		\exp \biggl(
			\alpha \delta_{r, r'}
			\sum_{p > 0} \frac{2\pi}{Lp}
			\ee^{\ii rp(x + \ii r 0^{+}) - \ii \omega(p)t}
		\biggr), \nonumber \\
R_{r, r'}(t)
& = L^{- \alpha \delta_{r, -r'}} 
		\exp \biggl(
			\alpha \sum_{p > 0} \frac{2\pi}{Lp}
			\frac{
				\ee^{-\ii\omega(p)t}
				\bigl(
					\delta_{r, r'} [\omega(p) - p \vF]^{2}
					+ \delta_{r, -r'} m^{2} \vF^4
				\bigr)
				- [\omega(p) - p \vF]^{2}
			}{2\omega(p) p \vF}
		\biggr), \nonumber \\
K_{r, r'}(x; t)
& = \sum_{p >0} \frac{2\pi}{Lp}\ee^{-\ii \omega(p)t} [\cos (px) - 1]
		\frac{
			\delta_{r, r'} [\omega(p) - p\vF]^{2}
			+ \delta_{r, -r'} m^{2} \vF^4
		}{\omega(p) p \vF}.	\label{AS_psipsi_2pcf_result_parts}
\end{align}
\end{widetext}
This result agrees with the two-point correlation functions for the (fermionic) Luttinger-Schwinger model in \cite{GLR}.
Moreover, $\tilde{G}_{r, r}(x; 0)$ coincides with the two-point correlation functions for free anyons in \eqref{Free_psipsi_2pcf_def} at equal times, while $\tilde{G}_{r, -r}(x; t) = 1$.
The zero modes would require a more detailed discussion which we again omit for conciseness.
We only note that they can be shown to give rise to the phase
$\ee^{\ii (r-r') \tilde{\theta}/2}$ in \eqref{AS_psipsi_2pcf_result}, which is essentially their only contribution that remains since we state the result in the limit $L \to \io$.

The correlation functions in \eqref{AS_psipsi_2pcf_result} and \eqref{AS_psipsi_2pcf_result_parts} have the following properties (all constants are $L$ independent):%
\begin{enumerate}[wide, labelindent=1em, nosep, label={\textnormal{(\roman*)}}, ref={\textnormal{(\roman*)}}]

\item
\label{No_ODLRO_AS:i}
If $m = 0$, then $K_{r, r'}(x; 0) = 0$ for $x \in \mathbb{R}$.
However, if $m > 0$, then there exist constants $C,C' > 0$ such that
\begin{equation}
\label{K_exponential_bound}
K_{r, r'}(x; 0) 
\leq - C m\vF |x| + \delta_{r, r'} C' (m\vF + \max \{ 1, \ln x^{2} \})
\end{equation}
for $|x| \geq (m\vF)^{-1}$.
Moreover, $\lim_{x \to 0} K_{r, r'}(x; 0) = 0$ for any $m\geq 0$.

\item
\label{No_ODLRO_AS:ii}
If $m = 0$, then
$R_{r, r'}(t) = L^{-\alpha \delta_{r, -r'}}$
for $t \in \mathbb{R}$.
However, if $m > 0$, then for $L > 2\pi$ there exist constants $C \in (0,1)$ (depending on both $m$ and $\vF$) and $c_{j}, C_{j} > 0$ for $j \in \mathbb{Z}$ (independent of both $m$ and $\vF$) such that
\begin{gather}
C \leq R_{r, r'}(0) \leq C^{-1}, \nonumber \\
R_{r, r'}(0)^{-1} R_{r, r'}(2\pi j / m\vF^{2}) \in [c_{j}, C_{j}]
\quad
\forall j \neq 0. \label{bounds_mathcalR0}
\end{gather}
Moreover,
$\lim_{L \to \io} R_{r,-r}(0) = \bigl( m\vF \ee^{\gamma} / 4\pi \bigr)^{\alpha}$ for any $m \geq 0$, where $\gamma$ is the Euler-Mascheroni constant.

\item
\label{No_ODLRO_AS:iii}
For $m > 0$, let $t \in [-\pi / m\vF^{2}, \pi / m\vF^{2}]$ and $j \in \mathbb{Z}$.
Then, for $L$ sufficiently large and for some constant $C >0$ (both depending on $t$, $j$, $m$, and $\vF$) the following bound holds:%
\begin{equation}
\bigl| R_{r, r'}(2\pi j / m\vF^{2} + t) \bigr| 
\leq \exp \bigl( -C t^{2} L \bigr).
\end{equation}

\end{enumerate}

Below, we only prove \ref{No_ODLRO_AS:i} and \ref{No_ODLRO_AS:ii}; the proof of \ref{No_ODLRO_AS:iii} is omitted since it is a straightforward extension of the one for \ref{No_ODLRO_AS:ii}.

\begin{proof}[Proof of \ref{No_ODLRO_AS:i}]
The fact that $K_{r, r'}(x; 0) = 0$ for $m = 0$ is obvious.
For $m >0$, note that
\begin{equation}
\label{manipulation_2pcf}
\frac{[\omega(p) - p\vF]^{2}}{\vF p^{2} \omega(p)}
= \frac{m^{2}\vF^{3}}{p^{2} \omega(p)}
	- 2 \biggl[ \frac{1}{p} - \frac{\vF}{\omega(p)} \biggr].
\end{equation}
Thus, we can write $K_{r, r'}(x; 0) = - 2f_{1}(x) + 2\delta_{r, r'} f_{2}(x)$ with
\begin{subequations}
\begin{align}
f_{1}(x)
& = \sum_{p > 0} \frac{\pi [1 - \cos(px)]}{L}
		\frac{m^{2} \vF^{3}}{p^{2} \omega(p)}, \label{f1_x} \\
f_{2}(x)
& = \sum_{p > 0} \frac{2\pi [1 - \cos(px)]}{L}
		\biggl[ \frac{1}{p} - \frac{\vF}{\omega(p)} \biggr]. \label{f2_x}
\end{align}
\end{subequations}
Since the functions above are even in $x$ we will henceforth restrict to $x \geq 0$.
We first obtain a lower bound for $f_{1}(x)$.
Note that it is a sum of positive terms and that
$m\vF^{2} \leq \omega(p) \leq m\vF^{2} + p \vF$ for $ p > 0 $. 
Thus, using $1 - \cos(px) \geq p^{2}x^{2}/8$ for $p \leq \pi/x$, we obtain
\begin{align}
\label{f_1_lower_bound}
f_{1}(x)
& \geq
	\frac{m^{2} \vF^{3}}{2(m\vF^{2} + \vF/x)}
	\sum_{0 < p \leq \pi/x} \frac{2\pi [1 - \cos(px)]}{L p^{2}} \nonumber \\
& \geq
	\frac{(m\vF x)^{2}}{16 (1 + m\vF x)}.
\end{align}
Using instead $1 - \cos(px) \leq p^{2}x^{2}/2$ for $p \leq m\vF$ and $1 - \cos(px) \leq 2$ otherwise, we obtain
\begin{align}
\label{f_1_upper_bound}
f_{1}(x)
& \leq
	\sum_{0 < p \leq m\vF} \frac{\pi x^{2} m\vF}{2L} 
	+ \sum_{p > m\vF} \frac{2\pi m\vF}{L p^{2}} \nonumber \\
& \leq (m\vF x/2)^{2} + m\vF \tilde{C}_{1}
\end{align}
for some constant $\tilde{C}_{1}$.
Consequently, we can take the limit inside the sum in \eqref{f1_x}, implying $\lim_{x \to 0} f_{1}(x) = 0$ by the dominated convergence theorem.
Next, we prove the following bound for $f_{2}(x)$:%
\begin{equation}
\label{f_2_upper_bound}
f_{2}(x) \leq \frac{C'}{2} \bigl( m\vF + \max \{ 1, \ln x^{2} \} \bigr)
\end{equation}
for some constant $C'$.
First, we consider $x \leq 1$, and we split the sum in \eqref{f2_x} into two parts, corresponding to the sums over $0 < p \leq m\vF$ and $p > m\vF$, respectively.
In the former we use
$1 - \cos p x \leq p x$ and $1/p - \vF/\omega(p) \leq 1 / p$.
In the latter we instead use $1 - \cos p x \leq 2$ and
$1/p - \vF/\omega(p) \leq m\vF / p^2$.
This implies
\begin{equation}
\label{f_2_small_bound}
f_{2}(x) \leq m\vF (x + \tilde{C}_{2})
\end{equation}
for some constant $\tilde{C}_{2}$.
By the same argument as below \eqref{f_1_upper_bound}, we have
$\lim_{x \to 0} f_{2}(x) = 0$.
We observe that this together with the corresponding result for $f_{1}(x)$ proves
$\lim_{x \to 0} K_{r, r'}(x; 0) = 0$.
Second, for $x > 1$, we split the sum in \eqref{f2_x} into three pieces:
$S_{1}$, $S_{2}$, and $S_{3}$,
corresponding to the sums over
$0 < p \leq x^{-1}$, $x^{-1} < p \leq 1$, and $p > 1$,
respectively.
Using $1 - \cos(px) \leq px$, we bound the first sum by
\begin{equation}
S_{1}
\leq \sum_{0 < p \leq x^{-1}} \frac{2\pi [1 - \cos(px)]}{Lp}
\leq \tilde{C}_{3}
\end{equation}
for some constant $\tilde{C}_{3}$.
We bound the second sum by
\begin{align}
S_{2}
\leq \frac{4\pi x}{L} + 2 \int_{x^{-1}}^{1} \frac{\dd p}{p}
\leq 2\pi + 2 \ln x.
\end{align}
Here, we used $1 - \cos(px) \leq 2$ to obtain a sum of a decreasing function, and then bounded that sum by the corresponding integral plus the first term in the sum.
(The bound is not optimal but sufficient for our purposes.)
Finally, we bound the third sum by again using
$1/p - \vF/\omega(p) \leq m\vF / p^{2}$
to obtain
\begin{equation}
S_{3}
\leq \sum_{p >1}
\frac{4\pi m\vF}{Lp^{2}} \leq m\vF \tilde{C}_{4}
\end{equation}
for some constant $\tilde{C}_{4}$.
Combining the contributions for $x \leq 1$ and $x >1$ implies \eqref{f_2_upper_bound}.
To conclude, \eqref{f_1_lower_bound} and \eqref{f_2_upper_bound} imply \eqref{K_exponential_bound} for $x \geq (m\vF)^{-1}$.
\end{proof}
\begin{proof}[Proof of \ref{No_ODLRO_AS:ii}]
By using \eqref{manipulation_2pcf} we can write
\begin{equation}
\bigl| R_{r, r'}(t) \bigr|
= R_{r, r'}(0)
	\ee^{ -\alpha \tilde{f}_{1}(t) + \alpha \delta_{r, r'} \tilde{f}_{2}(t) },
\end{equation}
where
\begin{equation}
\label{def_R_0}
R_{r, r'}(0)
= \exp
	\bigl(
		\alpha \delta_{r, -r'} [ -\ln L + S ]
	\bigr),
\end{equation}
with $S = \sum_{p > 0} (2\pi/L) \bigl[ 1/p - \vF/\omega(p) \bigr]$ and
\begin{subequations}
\begin{align}
\tilde{f}_{1}(t)
& = \sum_{p > 0} \frac{\pi [1 - \cos(\omega(p)t)]}{L}
		\frac{m^{2} \vF^{3}}{p^{2} \omega(p)}, \\
\tilde{f}_{2}(t)
& = \sum_{p > 0} \frac{2\pi [1 - \cos(\omega(p)t)]}{L}
		\biggl[ \frac{1}{p} - \frac{\vF}{\omega(p)} \biggr].
\end{align}
\end{subequations}
To begin, we obtain upper and lower bounds for $R_{r, r'}(0)$ that do not depend on $L$.
For some constants $\tilde{C}_{1}$ and $\tilde{C}_{2}$ depending on $m$ and $\vF$ only, the sum $S$ in \eqref{def_R_0} is bounded from below and above as follows:%
\begin{align}
S
& \geq \int_{2\pi / L}^{1} \dd p\, \biggl[ \frac{1}{p} - \frac{1}{m\vF} \biggr]
	\geq \tilde{C}_{1} + \ln L, \nonumber \\
S
& \leq 1 + \int_{2\pi/L}^{1} \frac{\dd p}{p} + \sum_{p > 1} \frac{2\pi m\vF}{Lp^{2}}
	\leq \tilde{C}_{2} + \ln L.
\end{align}
Since $1/p$ is a decreasing function, the lower bound was obtained by replacing the series with the integral, whereas the upper bound was obtained by furthermore adding the first term of the series.
This proves the first equation in \eqref{bounds_mathcalR0} uniformly in $L$.
For general $t \in (2\pi / m\vF^{2}) \mathbb{Z}$, we need to provide suitable bounds for
$\tilde{f}_{1}$ and $\tilde{f}_{2}$ uniformly in $L$.
By inspection, $\tilde{f}_{1}(2\pi j / m\vF^{2}) \geq 0$, while an upper bound is obtained as follows:%
\begin{align}
\tilde{f}_{1} \Bigl( \frac{2\pi j}{m\vF^2} \Bigr) 
& \leq
	\!\!
	\sum_{0 < p \leq m\vF}
	\!\!
	\frac{\pi \bigl[ 1 - \cos(2\pi j \sqrt{1 + p^2 / m^2\vF^2}) \bigr]}{L p^2 / m\vF}
	\nonumber \\
& \quad \;
	+ \sum_{p > m\vF} \frac{2\pi m\vF}{L p^{2}}
	\leq \tilde{c}_{j}
\end{align}
for some constants $\tilde{c}_{j}$, where we used that $\omega(p) \geq m\vF^2$ and that the function of $p$ in the first sum is continuous and bounded by $4\pi |j| / L m\vF$ on $[0, m\vF]$.
We also have $\tilde{f}_{2}(2\pi j / m\vF^{2}) \geq 0$, and an upper bound can be found similar to before, giving $\tilde{f}_{2}(2\pi j / m\vF^{2}) \leq 2 \tilde{c}_{j}$, where we instead used that $1/p - \vF/\omega(p) \leq m\vF / p^{2}$.
The above implies \eqref{bounds_mathcalR0}.
Finally, the limit below \eqref{bounds_mathcalR0} can be computed by noting that
\begin{align}
& \lim_{L \to \io}
	- \ln L
	+ \sum_{p >0} \frac{2\pi}{L}
		\biggl[\frac{1}{p} - \frac{\vF}{\omega(p)} \biggr]
	\nonumber \\
& \qquad
= \gamma
	+ \ln m\vF
	+ \lim_{M \to \io}
		\Biggl(
			\int_{2\pi}^{M} \frac{\dd \xi}{\xi}
			- \int_{0}^{M} \frac{\dd \xi}{\sqrt{\xi^2 + 1}}
		\Biggr),
\end{align}
where
$\gamma = \lim_{N \to \io} \bigl( - \ln N + \sum_{n = 1}^{N} n^{-1} \bigr)$.
\end{proof}
%


\section{Gauge field}
\label{App:Gauge_field}


The quantum fields $A(x)$ and $E(x)$, see Sec.~\ref{Sec:Extensions:AS_model}, formally satisfy $[A(x), E(x')] = \ii \delta(x-x')$.
We construct $\cF_{\mathrm{gauge}}$ by introducing boson creation and annihilation operators 
$a^{\pm}_{p}$ and a vacuum state $|\Psi_{\mathrm{gauge}, 0}\rangle$ such that
\begin{equation}
\bigl[ a^-_{p\ppr}, a^+_{p'} \bigr] = \d_{p, p'},
\quad
\bigl[ a^\pm_{p\ppr}, a^\pm_{p'} \bigr] = 0,
\quad
a^-_{p} |\Psi_{\mathrm{gauge}, 0}\rangle = 0
\quad
\forall p.
\end{equation}
We set
$\hat{A}(p) = (\ii / 2\pi) \sqrt{L/2} \bigl( a_{-p}^{-} - a^{+}_{p} \bigr)$
and
$\hat{E}(p) = \sqrt{L/2} \bigl( a_{-p} ^{-} + a^{+}_{p} \bigr)$
for $p \in (2\pi/L) \mathbb{Z}$.
It follows that
\begin{align}
\bigl[ \hat{A}(p), \hat{E}(p') \bigr]
& = \ii \frac{L}{2\pi} \delta_{p+p', 0}, \nonumber \\
\bigl[ \hat{A}(p), \hat{A}(p') \bigr]
& = \bigl[ \hat{E}(p), \hat{E}(p') \bigr]
= 0,
\end{align}
and
$\hat{A}(p) = \hat{A}(-p)^{\dagger}$,
$\hat{E}(p) = \hat{E}(-p)^{\dagger}$.
Finally, we formally write
$A(x) = \sum_{p} (2\pi/L) \hat{A}(p) \ee^{\ii p x}$ and
$E(x) = \sum_{p} (1/L) \hat{E}(p) \ee^{\ii p x}$.
[Note the extra factor $2\pi$ for $A(x)$.]


\subsection{Gauge transformations}
\label{App:Gauge_field:Gauge_transformations}


Gauge transformations are given by smooth periodic maps
$[-L/2,L/2] \to \mathrm{U}(1)$:
\begin{equation}
x \mapsto \ee^{\ii \Lambda(x)},
\quad
\Lambda(x) = \Lambda_{\text{large}}(x) + \Lambda_{\mathrm{small}}(x),
\end{equation}
where
$\Lambda_{\mathrm{large}}(x) = 2\pi w x/L$ for some $w \in \mathbb{Z}$
and
$\Lambda_{\mathrm{small}}(x)$ is a sufficiently smooth real-valued periodic function.
This gives a decomposition into large and small gauge transformations.

The gauge transformations act as follows:
\begin{align}
\psi^{q}_{r}(x; \eps)
& \mapsto
	\ee^{-\ii q \sqrt{\alpha\alpha_{0}}
	(\delta_{\eps} * \Lambda)(x)} \psi^{q}_{r}(x; \eps),
	\nonumber \\
\hat{A}(p)
& \mapsto
	\hat{A}(p) - \frac{ \sqrt{\alpha\alpha_{0}}}{2\pi e} \widehat{\partial \Lambda}(p),
	\nonumber \\
\hat{E}(p)
& \mapsto
	\hat{E}(p), \label{gauge_transf_action}
\end{align} 
where
$(\delta_{\eps} * \Lambda)(x)
= 2\pi w x/L
	+ \sum_{p} (1/L) \hat{\Lambda}_{\mathrm{small}}(p) \ee^{\ii p x - \eps |p|/2}$
and
$\widehat{\partial\Lambda}(p)
= \int_{-L/2}^{L/2} \dd x\, \partial_{x} [\Lambda(x)] \ee^{-\ii px}$.
Their implementers can be written as
$\mathcal{U}(\Lambda)
= \big(\mathcal{U}_{0}\big)^{w} \mathcal{U}(\Lambda_{\mathrm{small}})$
with
\begin{align}
\mathcal{U}_{0}
& = \ee^{\ii \pi \sqrt{\alpha_{0}} \rho(0)}
		\ee^{\ii \frac{2\pi}{L} \frac{\sqrt{\alpha\alpha_{0}}}{e} \hat{E}(0)}
		R_{+} R_{-},
		\nonumber\\
\mathcal{U}(\Lambda_{\mathrm{small}})
& = \exp \biggl(
			\ii \frac{\sqrt{\alpha\alpha_{0}}}{e}
			\sum_{p} \frac{1}{L} \hat{G}(p) \hat{\Lambda}_{\mathrm{small}}(-p)
		\biggr). \label{gauge_transf_l_s_impl}
\end{align}
Note that $\ee^{\ii \pi \sqrt{\alpha_{0}}\rho(0)}$ in $\mathcal{U}_{0}$ is needed to compensate for the phase factor due to the exchange relations of the Klein factors [cf.\ \eqref{R_exchange_relations}].

All physical quantities must be gauge invariant.
However, due to the nontrivial commutation relations in \eqref{rho_rho_CCR}
the densities are not:%
\begin{equation}
\mathcal{U}(\Lambda)^{-1} \rho_{r}(p) \hspace{0.3mm} \mathcal{U}(\Lambda)
= \rho_{r}(p) + r \frac{\sqrt{\alpha_{0}}}{2\pi} \widehat{\partial \Lambda}(p)
\quad \forall p.
\end{equation}
By comparing these transformations with those of the gauge field in \eqref{gauge_transf_action}, we observe that we can obtain gauge-invariant densities as
\begin{equation}
\label{def_gauge_invariant_densitites}
\tilde{\rho}_{r}(p)
= \rho_{r}(p) + r \frac{e}{\sqrt{\alpha}} \hat{A}(p)
\quad \forall p.
\end{equation}
Also, replacing
$\psi_{r}^{\mp}(x; \eps)$ by $\ee^{\pm \ii w_{\eps}(x)} \psi_{r}^{\mp}(x; \eps)$
with $w_{\eps}(x) = w_{\eps}(x, 0)$ in \eqref{Wilson-line_operator},
the anyonic fields can be made invariant under all nonrigid gauge transformations
[$\hat{\Lambda}_{\mathrm{small}}(0) = 0$].
These fields have a simple representation in terms of vertex operators:%
\begin{multline}
\label{gi_vertex_op}
\ee^{-\ii q w_{\eps}(x)} \psi_{r}^{q}(x; \eps)
= L^{-\alpha/2}
	\xxa
		R^{qr n_{\a}}_{r}
		\exp \biggl(
			- \ii qr \sqrt{\alpha} \\
\times
				\frac{2\pi}{L}
				\biggl[
					x \tilde{\rho}_{r}(0)
						+ \sum_{p \neq 0} \frac{1}{\ii p} \tilde{\rho}_{r}(p) \ee^{\ii px - \eps |p|/2}
				\biggr]
		\biggr) 
	\xxe
\end{multline} 
for $q = \pm$, which is used for actual computations.
In Appendix~\ref{App:Gauge_field:Anyon_normal_ordering} we will show that the definition in \eqref{def_gauge_invariant_densitites} is natural and equivalent to introducing Wilson lines.


\subsection{Gauss's law and physical Hilbert space}
\label{App:Gauge_field:Gauss_law}


The physical Hilbert space $\cH_{\mathrm{phys}}$
is introduced by fixing the action of gauge transformations.
In what follows we assume that an external charge density $\rho_{\mathrm{ext}}(x)$ is present and let $\rho_{\mathrm{ext}}(p) = \int_{-L/2}^{L/2} \dd x\, \rho_{\mathrm{ext}}(x) \ee^{-\ii px}$.

Let $|f\rangle$ be any state in $\cH_{\mathrm{phys}}$.
The action of small gauge transformations is fixed by imposing Gauss's law [cf.\ \eqref{Gauss_law_ext_charges}]
\begin{equation}
\hat{G}(p) |f\rangle
= - \frac{e}{\sqrt{\alpha}} \rho_{\mathrm{ext}}(p) |f\rangle
\end{equation}
so that we have the following $U(1)$-representation:%
\begin{equation}
\mathcal{U}(\Lambda_{\mathrm{small}}) |f\rangle
= \ee^{-\ii \sqrt{\alpha_{0}}
	\sum_{p} L^{-1} \rho_{\mathrm{ext}}(p) \hat{\Lambda}_{\mathrm{small}}(-p)} |f\rangle,
\end{equation}
provided that \hspace{-0.9mm} $\Lambda_{\mathrm{small}}$ is regular enough,
\hspace{-0.8mm} depending on \hspace{-0.8mm} $\rho_{\mathrm{ext}}(\hspace{-0.1mm}x\hspace{-0.2mm})\hspace{-0.1mm}$. 
On the other hand, large gauge transformations are fixed by requiring that
\begin{equation}
\mathcal{U}_{0} |f\rangle = \ee^{\ii \theta} |f\rangle
\end{equation}
for some fixed $\theta \in [0,2\pi)$.
The dependence of the representation on a ``free'' parameter $\theta$ is a consequence of the global chiral symmetry breaking: 
The generator of global chiral transformations
$\rho_{5}(p = 0) = \sqrt{\alpha_{0}}(Q_{+} - Q_{-})$
is not gauge invariant, and thus chiral transformations cannot be represented in $\cH_{\mathrm{phys}}$, i.e., they are broken and a continuous set of gauge-invariant vacua is possible \cite{Stro}.
In general, $\cH_{\mathrm{phys}}$ thus depends on the set
$\{ \theta; ( \rho_{\mathrm{ext}}(p) )_{p \in (2\pi/L)\mathbb{Z}} \}$,
but this dependence is usually attached only to the vacuum;
the standard notation $|\Psi_{\mathrm{AS}}(\theta)\rangle$ is adopted when $\rho_{\mathrm{ext}}(x) = 0$.


\subsection{Gauge-invariant anyon normal ordering}
\label{App:Gauge_field:Anyon_normal_ordering}


Anyon normal ordering can be made gauge invariant by inserting a Wilson line
$W_{\eps}(x, x') = \ee^{-\ii [w_{\eps}(x) - w_{\eps}(x')]}$
between the fields [see also \eqref{AS_psipsi_2pcf_def}]:
\begin{equation}
\psi^{+}_{r}(x; \eps) \psi^{-}_{r}(x'; \eps)
\mapsto
\psi^{+}_{r}(x; \eps) W_{\eps}(x, x') \psi^{-}_{r}(x'; \eps)
\end{equation}
and
\begin{align}
& \psi^{+}_{r}(x; \eps)
	[ - \ii \partial_{x'} + e A(x') ]
	\psi^{-}_{r}(x'; \eps)
	\nonumber \\
& \quad
\mapsto
\psi^{+}_{r}(x; \eps)
	W_{\eps}(x, x') [ - \ii \partial_{x'} + e A(x') ]
	\psi^{-}_{r}(x'; \eps).
\end{align}
We therefore formally define
\begin{multline}
\anyonWick{
	\psi^{+}_{r}(x) \partial^{n}_{x} \psi^{-}_{r}(x)
}{\mathrm{AS}} \\
= \lim_{\eps \to 0^{+}} \lim_{y \to 0}
	\anyonWick{
		\psi^{+}_{r}(x; \eps) \partial^{n}_{y} \psi^{-}_{r}(x+y; \eps)
	}{\mathrm{AS}, \eps}
\end{multline}
for $n = 0, 1, \ldots$, where
\begin{multline}
\anyonWick{
	\psi^{+}_{r}(x; \eps) \partial_{y}^{n} \psi^{-}_{r}(x+y; \eps)
}{\mathrm{AS}, \eps}
= \frac{r}{2\pi \ii} \frac{L^{\alpha}}{(n+1)!} \\
\times
	\biggl( \frac{1}{\sqrt{\alpha}}\frac{\partial}{\partial y} \biggr)^{n+1}
	\! \bosonWick{ \psi^{+}_{r}(x; \eps) W_{\eps}(x, x+y) \psi^{-}_{r}(x+y; \eps) } \!
\end{multline}
for $y \neq 0$.
By standard manipulations with vertex operators,
\begin{multline}
\! \bosonWick{ \psi^{+}_{r}(x; \eps) W_{\eps}(x, x+y) \psi^{-}_{r}(x+y; \eps) } \!
= L^{-\alpha}
	\! \xxa
		\exp \biggl(
			\ii r \sqrt{\alpha} \\
\times
			\frac{2\pi}{L} \biggl[
				y \tilde{\rho}_{r}(0)
				+ \sum_{p \neq 0} \frac{\ee^{\ii p y} - 1}{\ii p}
					\tilde{\rho}_{r}(p)
					\ee^{\ii p x - \eps |p|/2}
			\biggr]
		\biggr)
	\xxe \,,
\end{multline}
using $\tilde{\rho}_{r}(p)$ in \eqref{def_gauge_invariant_densitites}.
As in Appendix~\ref{App:Anyon_normal_ordering}, one can show that 
$\anyonWick{ \psi^{+}_{r}(x)\psi^{-}_{r}(x) }{\mathrm{AS}}$ can formally be identified with 
$\tilde{\rho}_{r}(x)$ and
\begin{multline}
\int_{-L/2}^{L/2} \dd x\,
	\anyonWick{
		\psi^{+}_{r}(x) r\vF [-\ii \partial_{x} + e A(x)] \psi^{-}_{r}(x)
	}{\mathrm{AS}} \\
=	\frac{\pi\vF}{L} \tilde{\rho}_{r}(0)^2
	+ \sum_{p>0} \frac{2\pi\vF}{L} \tilde{\rho}_{r}(-rp)\tilde{\rho}_{r}(rp)
\end{multline}
for $r = \pm$, which shows that the formal Hamiltonian in \eqref{H_AS} can be written using densities [cf.\ \eqref{H_AS_Fourier_space}].


\end{appendix}



\end{document}